%% file: main.tex
\documentclass[aoas]{imsart}

\input{myprem}

\usepackage{comment}
\usepackage{soul}
\usepackage{xcolor}
\def\mo#1{\textcolor{black}{#1}}
\def\lc#1{\textcolor{black}{#1}}

\startlocaldefs

\endlocaldefs

\begin{document}
\graphicspath{{img/}}
\begin{frontmatter}
  
\begin{aug}
\title{Predictive inference for travel time on transportation networks}
\runtitle{Predictive inference for travel time on transportation networks}
 
\author[A]{\fnms{Mohamad} \snm{Elmasri}\ead[label=e1]{mohamad.elmasri@utoronto.ca}},
\author[B]{\fnms{Aur\'elie} \snm{Labbe}\ead[label=e2]{aurelie.labbe@hec.ca}},
\author[B]{\fnms{Denis} \snm{Larocque}\ead[label=e3]{denis.larocque@hec.ca}} and
\author[B]{\fnms{Laurent} \snm{Charlin}\ead[label=e4]{laurent.charlin@hec.ca}}

\address[A]{Department of Statistical Sciences, University of Toronto, \printead{e1}}
\address[B]{Department of Decision Sciences, HEC Montr\'eal, \printead{e2,e3,e4}}
\end{aug}

\begin{abstract}
Recent statistical methods fitted on large-scale GPS data \lc{can provide accurate estimations of the expected travel time between two points.}
{However, little is known about the distribution of travel time, which} is key to decision-making \lc{across a number of logistic problems}. { With sufficient data, single road-segment travel time can be well approximated}. The challenge lies in understanding how to aggregate such information over a route to arrive at the route-distribution of travel time. We develop a novel statistical approach to this problem. We show that, under general conditions, without assuming a distribution of speed, travel time {divided by route distance follows a Gaussian distribution} with route-invariant population mean and variance. We develop efficient inference methods for such parameters and propose asymptotically tight population prediction intervals for travel time. \mo{Using traffic flow information}, we further develop a trip-specific Gaussian-based predictive distribution, resulting in tight prediction intervals for short and long trips. Our methods, implemented in an R-package\footnote{Available at \texttt{https://github.com/melmasri/traveltimeCLT}.}, are illustrated in a real-world case study using mobile GPS data, showing that our trip-specific and population intervals both achieve the 95\% theoretical coverage levels. Compared to alternative approaches, our trip-specific predictive distribution achieves (a) the theoretical coverage at every level of significance, (b) tighter prediction intervals, (c) less predictive bias, and (d) more efficient estimation and prediction procedures. This makes our approach promising for low-latency, large-scale transportation applications.
  \end{abstract}

\end{frontmatter}
\tableofcontents
\section{Introduction}
\subsection{Background}
Mobility is vital to  human activities, as it is an integral component of our economic and trade networks, social interactions, political ties, and our quality of life. The growing population, as well as new transportation methods and systems, are posing challenges to our current transportation networks. Large-scale trip-level data, with temporal and spatial coverage, enable us to better diagnose transportation problems and develop effective solutions to such things as increasing congestion levels. Such data is made available progressively as it is collected, for example, from global positioning systems (GPS) on mobile phones and other devices, or from radar systems used in aerial and marine positioning.


{
Accurate estimation of travel time is critical for a wide range of transportation-related applications, such as online routing services, ride-share platforms\footnote{As examples, Google Maps, Lyft Inc., and Uber Inc.} , and freight and shipping services. These systems make millions of operational and pricing decisions based on travel time estimates, which require both a thorough understanding of the distribution of travel time and reliable methods for predicting various quantities related to this distribution. For example, estimates of the probability of delay can help matching systems dispatch rides more efficiently than simply dispatching the nearest ride (i.e., the one with the smallest expected travel time). To determine the number of rides that are within a five-minute window of a specific location with 95\% probability, we need to access the 95th percentile of the distribution of travel time. By using this information, we can find a better ride that allows sufficient time for a rider to be ready for the pickup while maximizing the overall utility of the system. Previous research has shown that incorporating uncertainty into travel time estimates can lead to improved outcomes, such as higher driver-rider match rates (up to 85\%) and lower average prices (up to 12\%) in ride-share systems~\citep{long2018ride}. These improvements can also impact other key system metrics, such as reliability, service rate, utilization, and profitability~\citep{li2021vehicle, li2022pricing,li2022ride}.}

\subsection{Motivating toy example}
{We first provide some intuition regarding the distribution of travel time in a simulated toy example. We generate several rides 
on the same 100-edge route, each edge of 100 meter length, 
in the following way. For every edge of every ride, we sample the edge speed from a log-normal mixture over two states, non-congested and congested, as  in~\citep{woodard2017predicting,guo2012multistate}. The non-congested state occurs with 0.8 probability, with a mean of 35~km/h and a standard deviation of 5~km/h, sampled as logNormal\((\mu=\ln(35),\sigma=\ln(5))\). The congested state has a mean of 5~km/h with 10~km/h as a standard deviation. We set the start time of each ride to 0. Two of these rides with a similar total travel time are selected. Their trajectories are shown in 
Figure~\ref{fig:filtration-two-trips} (left).
\begin{figure}[ht!]
  \centering
  \includegraphics[width=0.59\textwidth]{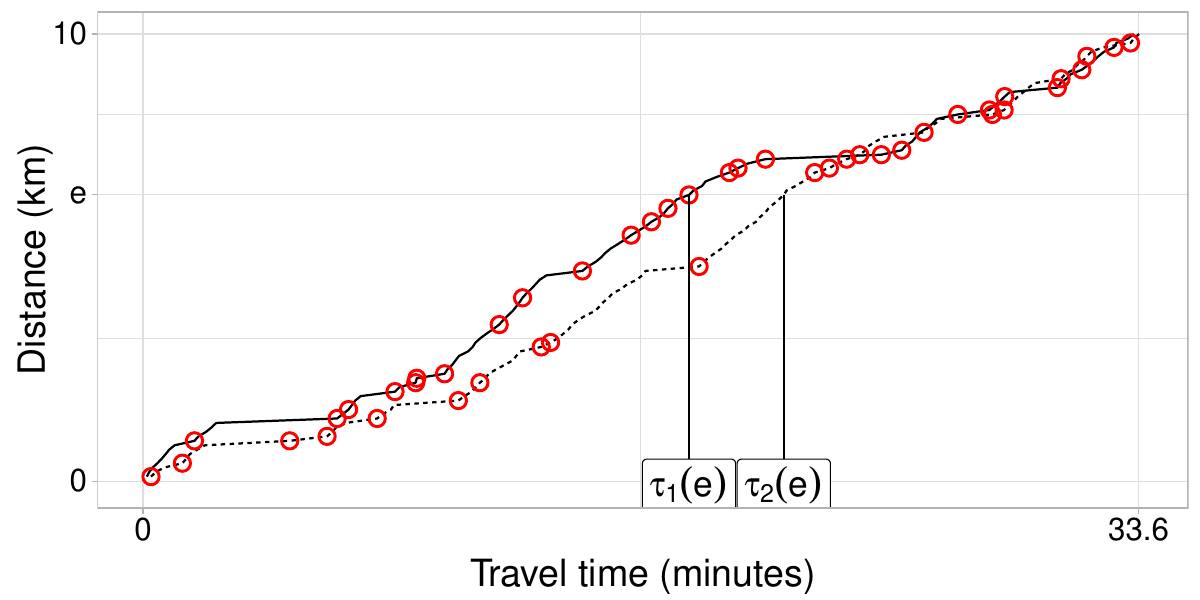}
  \includegraphics[width=0.40\textwidth]{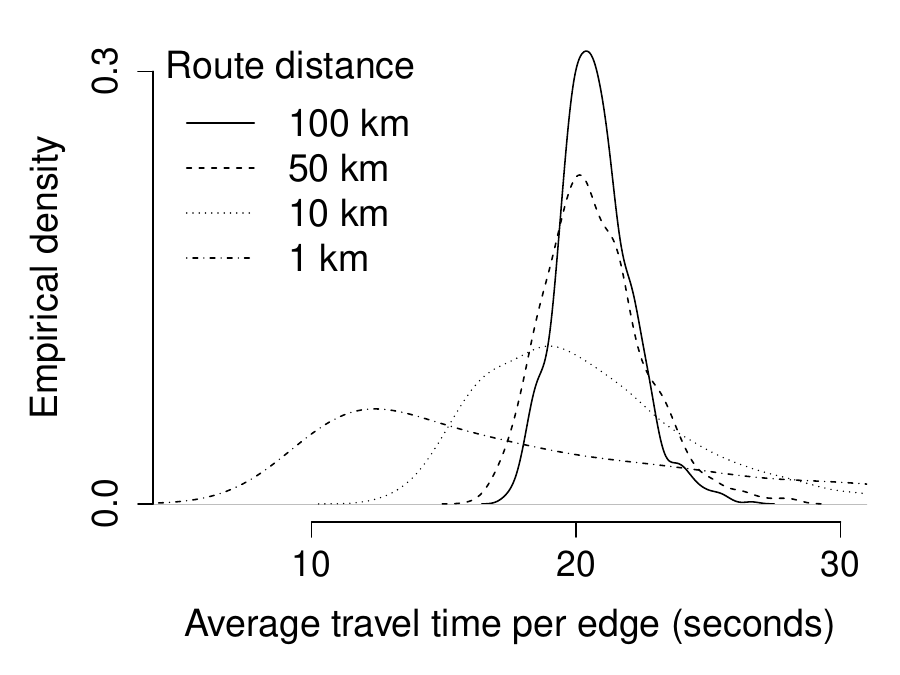}
  \caption{{ (Left) A simulated toy example involving two vehicles travelling a 100-edge route, of 100 meter edges, with \(\tau_{i}(e), i=1,2,\) being the vehicles' travel times up to \(e\). (Right) Empirical distribution of average travel time per edge from 1000 samples over a 1000-edge route.}}
    \label{fig:filtration-two-trips}
  \end{figure}

The circles (red) in the figure depict the congested states encountered by each vehicle. We observe that: 1) the travel time up to edge \(e\) is different for the two vehicles; and yet, 2) this does not imply that the long-term travel behaviour of the two vehicles is statistically different. In this example, the short-run travel time difference up to edge $e$ is caused by 14 congestion events encountered by vehicle 1, while vehicle 2 encounters only 10. However, such short-run difference corrects over the long-term, as shown by the two curves intersecting every now and then in this example.

  To illustrate the long-term behaviour, we repeat this sampling process for 1000 vehicles over a 1000-edge route totaling 100 km. Figure~\ref{fig:filtration-two-trips} (right) depicts the empirical density of the average travel time per edge for the first 1, 10, 50, and 100 km of the route. The average is calculated per vehicle and route distance as the total time divided by the number of edges of the route. The depicted density of the 1st km exhibits heavy skewness and large variability. As vehicles progress, they encounter more congestion events. However, the distribution of the average travel time per edge becomes more and more symmetric, around 21 seconds, with less variability.

  Such observation generalizes the two-vehicle example of  Figure~\ref{fig:filtration-two-trips} (left) to explain that: a) there is an asymptotic distribution for travel time over a route; and b) aggregating information from multiple trips (to estimate edge speed) can be used to approximate this distribution. 
}
\subsection{Objectives}
We consider a {\it transportation network} \(G\) to be a graph with stochastic edge-weights (speeds) governing the travel time of the edge. A {\it route} \(\path\) on $G$ consists of a \(n\)-sequence of connected edges $\path = \langle e_1, e_2, \dots, e_n\rangle$ that define the order of travel. Travel time over \(\path\) is defined as a random variable through the sum
\begin{equation}
  \label{eq:speed-sum-cond}
  \T_{\path} = \sum_{e \in \path} d_{e} S_{e}, 
\end{equation}
where $S_{e}$ is the speed over the \mo{edge $e$ of length \(d_{e}\)}. \mo{We measure speed, $S_e$, in units of time over distance (rather than distance over time), because it simplifies our derivations.}

In real-world road, the distribution of speed \(S_{e}\) in~\eqref{eq:speed-sum-cond} is well-studied empirically~\citep{woodard2017predicting,guo2012multistate}. However, the different types of dependencies among speeds of a trip render it difficult to infer the distribution of \(\T_{\path}\) from the distribution of its components in~\eqref{eq:speed-sum-cond}.

Our first objective is to characterize the long-term (asymptotic) distribution of travel time, in which we show that $\T_\rho$ divided by route distance is asymptotically Gaussian (Sec.~\ref{sec:asymptotic-properties}). To estimate the parameters of this distribution, we must take into account two kinds of temporal dependencies: 1) the \emph{within-trip dependency} between a ride's speed observations; and 2) the \emph{temporal (or filtration) dependency}, which implies that the speed distribution at an edge depends on the arrival time at that edge. {Not accounting for such dependencies} can yield erroneous estimates of the {expected value} of the variables in the sum in~\eqref{eq:speed-sum-cond}, which in turn leads to the propagation of predictive bias with distance. {By accounting for such dependencies}, we provide Gaussian-based prediction intervals for travel time that are wide in the short-term, but which converge to zero asymptotically {with respect to distance} (Sec.~\ref{sec:point-wise-asympt}).

Our second objective is to provide prediction intervals for travel time that are tight for short and long trips. We do so by constructing route-specific mean and covariance sequences that integrate traffic information on a route. Those sequences centre and normalize travel time to a Gaussian-based predictive distribution (Sec.~\ref{sec:predictive-distribution}). With the route-specific sequences, we form prediction intervals that are significantly tighter for short trips (almost by 50\% as shown in our case study, Sec~\ref{sec:pool-based-prediction-intervals}) than the asymptotic prediction intervals while maintaining the desired coverage level. We illustrate the various properties of our methods in a real-world case study using GPS data collected from mobile phones in Quebec City.

\subsection{Outline}
We begin by introducing the GPS trip data collected by users in Quebec City, Canada, which will be used for illustration purposes (Sec.~\ref{sec:data-intro}). We then characterize transportation networks \mo{as directed graphs} with stochastic edge-speeds (Sec.~\ref{sec:transportation-networks}) and introduce general forms of dependencies between sequences of random variables to capture spatial and temporal dependencies in travel time (Sec.~\ref{sec:travel-time-random-variable}). We build on the notion that random variables far apart in a sequence are nearly independent in order to define a form of temporal dependency that is more general than, and includes, the commonly used Markov dependency. This leads us to define travel time as a dependent sampling process on a network (Sec.~\ref{sec:travel-time-sampling}). With this definition and the assumption of temporal-cyclicality of speed, our first contribution shows that the mean and variance parameters of the asymptotic normal distribution are route and start-time invariant (Sec.~\ref{sec:asymptotic-properties}). Using this result, we develop inference methods for those parameters (Sec.~\ref{sec:estimation-mu-sigma}), provide an asymptotic Gaussian-based confidence interval for the mean (Sec.~\ref{sec:conf-bounds}) and asymptotic population prediction intervals for trips (Sec.~\ref{sec:point-wise-asympt}).

We focus on long-term (asymptotic) and short-term behaviours of travel time. By utilizing and extending some recent results in dynamical systems, our second contribution is to build trip-specific mean and covariance sequences, in Section \ref{sec:predictive-distribution}, that are (i) calculated at the start of the trip, (ii) can centre and normalize travel time to an asymptotic Gaussian-based predictive distribution, and (iii) attain 95\% predictive coverage across the whole trip length. The tightness of the bounds in (iii) depends solely on the number of higher-order auto-correlation parameters between route edges estimated and included in the trip-specific covariance. 

In Section~\ref{sec:case-study}, we study our method using the Quebec City data. We first show that, empirically, higher traffic density reduces the auto-correlation between the variables in~\eqref{eq:speed-sum-cond} while increasing travel time variability (Sec.~\ref{sec:parameter-estimation}), an expected phenomenon in empirical travel time. Later, we compare our asymptotic route-invariant prediction intervals to our trip-specific intervals to establish empirically that: 1) by including only the first-order auto-correlation, the trip-specific intervals are approximately half as tight as the asymptotic intervals while attaining theoretical coverage levels; and 2) that adding higher-order auto-correlation terms does not necessarily lead to significantly tighter intervals (Sec~\ref{sec:pool-based-prediction-intervals}). Lastly, to illustrate points (a)--(d) of the abstract, we compare our trip-specific intervals to competing models in Section.~\ref{sec:comp-altern-models}.

\subsection{Literature review}
\mo{ Interest in understanding travel uncertainty has been growing. Works in this area can be classified into three categories, those that model i) travel time of individual edges, ii) route and travel uncertainty simultaneously, and iii) route-conditional travel times. Models in category i) focus on providing accurate edge-level distribution of travel time~\citep{zheng2013urban,jenelius2013travel}}. Models in category ii) attempt to integrate Bayesian methods and/or assume a parametric route-invariant distribution for travel time, with log-normals being the most commonly used~\citep{westgate2013travel,wang2019simple, hunter2009path,westgate2016large}. With the increasing volume of collected GPS data, models in category iii) integrate edge-level traffic data over a whole route to arrive at route-conditional travel times~\citep{woodard2017predicting,guo2012multistate,ma2017estimation,zhang2019novel}. While promising~\citep{yan2021statistical}, most of the latter models either require strong assumptions or are strongly overparametrized machine learning methods that can generate unnecessary computational costs. 

A large amount of work exists that models {stochastic processes on networks}, in biology, social networks, and other network types. {Most of this work attempts to understand generic properties of network-related systems, under specific analytical models, but not travel time in particular}. For a survey of dynamical processes on networks (including transportation networks), see~\citet[ch. 11]{barrat2008dynamical}. The statistics community has also shown  interest in developing tools for network analysis \citep{kolaczyk2009models,kolaczyk2014statistical}, including some to model {stochastic} processes on networks \citep{ramsay2007parameter, snijders2017modeling,burk2007beyond, britton2002bayesian,golightly2005bayesian,bolin2022gaussian}. Our work falls into this category, by providing a limit theorem for a type of mixing process on ergodic dynamical networks, which allows for efficient statistical inference and predictive methods.

\section{Quebec City trip data}
\label{sec:data-intro}
\subsection{Data collection}
Quebec City 2014 GPS data (QCD) is collected using the \textit{Mon Trajet} smartphone application developed by Brisk Synergies Inc. This study uses a sample of open data,\footnote{{A cleaned sample of the data is available at \texttt{https://github.com/melmasri/traveltimeCLT}}.} which contained 21,872 individual trips. The sample contains no data that can be linked to individual drivers. While no data was collected during the winter months, the precise duration of the collection period is kept confidential. \mo{The application was installed voluntarily by over 4,000 drivers, who then anonymously logged information using a simple interface. The exact number of drivers is kept confidential.}

\subsection{Data cleaning}
No measure was provided to ensure the validity of trips, i.e.~if they were made solely by motor vehicles and not walkers or cyclists, and excluded non-traffic interruptions such as parking. Data is processed by breaking down trips into multiple trips whenever: i) trips include idle time (a period of no movement) of more than 4 minutes; or ii) there are more than 2 minutes between consecutive GPS observations. To remove warm-up and parking periods, the end-points of the decomposed trips are trimmed, such that each trip starts whenever the vehicle speed reaches greater than 10~km/h for the first time and ends whenever the speed is less than 10~km/h for the last time. To remove non-motorized travel, we filter out trips with a median speed of less than 20~km/h and a maximum speed of less than 35~km/h, or those whose driving distance is less than 1~km (as measured by the sum of the great-circle distances between pairs of sequential measurements). \mo{We focus on personal vehicular travel. Nonetheless, some trips may be from other transit modes, such as bikes or buses. This is an unavoidable challenge when using mobile GPS data.}

The cleaned data contains 19,967 trips, which are composed of a sequence of GPS readings. The total trip duration is the difference between the last and first GPS timestamps. The median and average trip durations are 19 and 21~min, respectively, with a maximum of 3~h~27~min. The median trip distance is 14.5~km, with an average of 16.6~km and a maximum of 170.4~km. The median time between consecutive GPS points is 4~s, and the average is 9~s.

\subsection{Map matching}
A third-party service (TraxMatching\footnote{{\texttt{https://www.traxmatching.io}}}) was used to map trips' GPS observations to the Quebec City road network mapped by The OpenStreetMap Project\footnote{\texttt{https://www.openstreetmap.org}} (OSM), a publicly accessible open-source project. This process is called map-matching, and numerous high-quality methods are available to do this \citep{newson2009hidden,hunter2013large}. For each trip, the third-party service returns a sequence of mapped GPS points with lengths equal to the original sequence. Each mapped GPS point is associated with a source ``node id'', ``way id'' and destination ``node id'' corresponding to a unique directional edge whose ``way id'' is between the source and destination nodes. The map-matching process resulted in 46,386 unique directional edges, which constitute the travelled portion of Quebec City, {not its entirety.} The average edge length is 170~meters, and the median is 81~m. 

\subsection{Traffic estimation}
We estimate the total travel time per edge by calculating i) within-edge travel time, as the time spent within the edge, and ii) across-edge travel time, as the time spent between the two closest GPS observations, where one is in the edge and the other is in an adjacent edge. We calculate the across-edge distance in the same way. The total travel time per edge is then 100\% of within-edge plus across-edge travel time, weighted by half the proportion of across-edge distance to the total length of the edge. Total edge lengths are obtained from OSM. In rare circumstances, the map-matching service also returns intermediate edges that do not have initial GPS observations. This happens, for example, when a vehicle is moving fast or through a tunnel. We treat those intermediate edges, those without GPS observations, as a single edge and calculate the total travel time over it, and then assign edge travel time proportionally to the length of each intermediate edge. With these total travel time estimates, we calculate the average speed per edge by dividing the total travel time by the total length.

\begin{figure}[!htbp]
  \centering
  \includegraphics[width=0.6\textwidth]{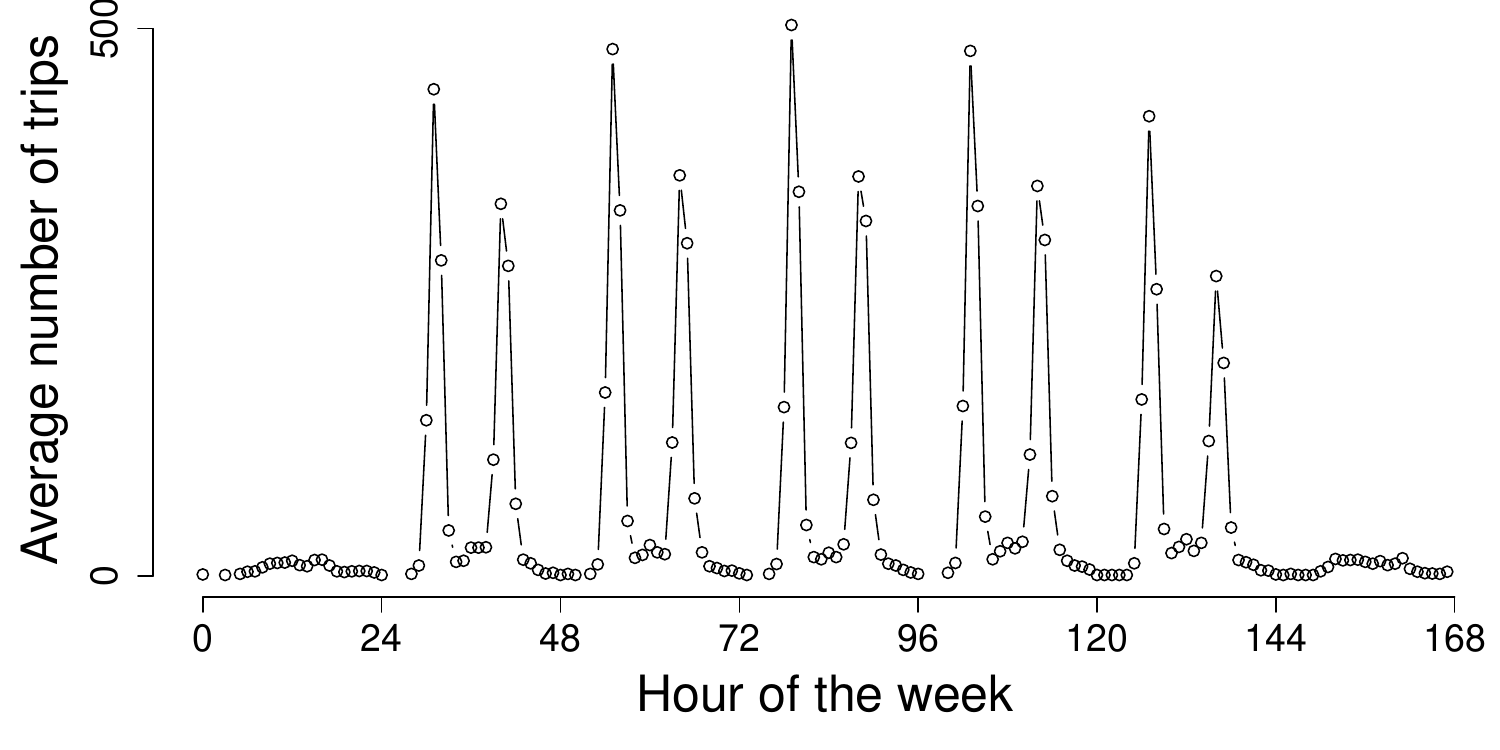}
  \caption{Average number of trips per hour in QCD, by hour of the week.}
  \label{fig:quebec-city-time-per-hour}
\end{figure}

Figure \ref{fig:quebec-city-time-per-hour} shows seasonal (weekly) traffic patterns per week hour, starting at the first hour of Sunday. The volume of traffic is reduced overnight on weekdays starting after 7~p.m.~and on weekends. Daily traffic peaks are associated with a.m.~and p.m.~rush hours, with strong dips in between.

\subsection{Data splitting} \label{subsec:split}
Because of the seasonal (weekly) traffic pattern illustrated in Figure~\ref{fig:quebec-city-time-per-hour} and the sparsity of GPS data for each edge, we classify our traffic data into three traffic time-bins (traffic-bins): i) an ``AM-rush hour'' bin for weekdays 6:30–8:30AM; ii) a ``PM-rush hour'' bin for weekdays 3:30–5:00~p.m., and iii) a ``Non-rush hour'' bin for all remaining periods. Alternative traffic-bins have been tested, but we found that aggregating data into these bins yielded the best results. In QCD, 37\% of trips occurred in an AM-rush hour, with the \mo{same proportion} for the PM-rush hour, and 26\% in all other periods.  

A test set of trips is sampled at random, but sequentially. That is, if removing a trip from the data results in an edge$\times$traffic-bin with \mo{no observations}, then the trip is not removed. After this process is carried out, 2,000 trips are set aside (851 AM-rush stratum, 741 PM-rush, and 408 Non-rush), with the remaining 17,967 trips being the training data. We classify a trip into a bin if all trip-edges are travelled within that bin; otherwise, the speed data for those trips are used for traffic estimation, but not trip analysis.
\section{Statistical framework}
\subsection{Transportation networks} \label{sec:transportation-networks}
\subsubsection{Network notations}
We define a transportation {network} \(G = (N, E, D, {S})\) as a directed connected graph consisting of a finite {\it node} set \(N\) and an {\it edge} set \(E\). For each edge \(e \in E\), \(d_{e} \in D\) defines the edge traversal positive distance, {and $S_e$ is the edge speed (defined in the next section)}. \(G\) is connected in the sense that a traversable route between any two nodes of $G$ exists. A {\it route} $\path$ in $G$ consists of a \(n\)-sequence of connected edges $\path = \langle e_1, e_2, \dots, e_n\rangle$ that define the order of travel. We {define} a route \(\path\) in \(G\) by the angle bracket \(\langle \cdot \rangle\), such that \(\langle e, e' \rangle \in \path\) is a subroute composed of a pair of edges \(e\) and \(e'\). {The notation} \(\langle \dots, e \rangle\) refers to a subroute in \(\path\), up to and including edge \(e\). Lastly, \(n = \#\{e: e\in \path\}\) refers to the length (number of edges) of \(\path\) and \(n_{e}\) to the number of times edge \(e\) appears in \(\path\). We assume that all edges of \(\path\) are fully travelled.

\subsubsection{Distribution of speed}
Let the continuous map \((s, t) \mapsto F_{e}(s, t) :=\P(S_e \leq s \mid t)\), in both \(s\) and \(t\), represent the cumulative distribution function (CDF) of the speed \(S_{e}\), {for a time \(t>0\), and value $s>0$.} The function \(F_{e}(\cdot, t)\) is a strictly increasing function. A random speed observation at time \(t_{0}\) is defined by its inverse CDF as \(S_{e}(t_{0}) = F^{-1}_{e}(U_e, t_{0})\), where \(U_e\) is a Uniform\([0,1]\) random variable and \(F^{-1}\) is the inverse CDF. The temporal distribution of speed on the network \(G\) can be represented as
\begin{equation}
\label{eq:cdf-speed-network}
 (S_{e}(t), e \in E) \stackrel{d}{=} (F^{-1}_{e}(U_{e},t), e\in E), \quad (U_{e}, e\in E) \stackrel{i.i.d}{\sim} \text{Uniform}[0,1],
\end{equation}
where \(\stackrel{d}{=}\) implies equality in distribution. {In many real-world transportation networks, speed is directly influenced by the traffic flow on the network. For example, during rush hours, average speeds tend to be lower on most edges \citep{treiber2000congested,williams2003modeling}. {This relationship has been studied extensively in what is known as network-wide fundamental diagrams in~\cite{geroliminis2008existence}.} Therefore, we assume a causal relationship between network traffic flow, what we term {\it traffic-states}, and individual edge speeds. To capture this relationship, we represent traffic-states in $G$ by $t\mapsto \Pi(t)$, a \mo{stochastic process} that is continuous with respect to $t$. Given $\Pi$, we assume that speeds on adjacent edges are conditionally independent, that is} 
\begin{equation}
    \label{eq:speed-regimes}
    S_{e} \independent \; S_{e'} \mid \Pi, \qquad \text{for all } e, e' \in E, e\neq e'.
\end{equation}

We further assume that \((S_{e}, e \in E)\) are strictly positive and bounded random variables, as they cannot be zero over an edge with positive length. We define \(S_{e}\) as follows.
\begin{definition} \label{assump:seasonality-of-speed-on-a-road} For a time index \(t >0 \), assume that the distribution of speed \(S_{e}(t)\) over an arbitrary edge \(e \in E\) is wide-sense cyclostationary, as
\begin{equation}
  \label{eq:seasonality-of-speed}
  S_{e}(t) =  m_{e}(t) + \epsilon_{e}(t),
\end{equation}
where {\(S_{e}(t) \in  [s_{\min{}} , s_{\max{}} ]\) for some absolute constants \(0<s_{\min{} }\leq s_{\max{} } <\infty\)}, \(m_{e}(t)=\E[ S_{e}(t)]\) and \(\sigma^{2}_{e}(t) =\E[\epsilon^{2}_{e}(t)] >0\) are continuous cyclostationary functions with respect to \(t\), with \(\E[\epsilon_{e}(t)]=0\) for all \(t\), such that $\epsilon_e (t) \independent m_e(t) \mid \Pi$.
\end{definition}

Wide-sense cyclostationarity is what is referred to as the cyclostationarity of \(m_{e}(t)\), and \(\sigma_{e}(t)\), having periodic and/or seasonal patterns~\citep{gardner2006cyclostationarity}. {It is possible that multiple periodic trends, for example, with weekly, daily, and/or hourly cycles. Those trends can be modelled additively through the mean}. Seasonality of \(m_{e}(t)\) is also justified by the periodic empirical behaviour of speed in real-world road networks, discussed in various forms \citep{williams2003modeling,jenelius2013travel, zheng2013urban, wang2019simple, woodard2017predicting}. {The constants $s_{\min{}}$ and $s_{\max{}}$ in Definition~\eqref{assump:seasonality-of-speed-on-a-road} represent the minimum and maximum speeds on the network. In the next section, we build on Definition~\ref{assump:seasonality-of-speed-on-a-road}, to define the distribution of travel time over a route.}

\subsection{Travel time as a random variable}
\subsubsection{Dependency in travel time}\label{sec:travel-time-random-variable} 
The main difficulty in inferring the distribution of $\T_{\path}$ is the different sources of dependencies affecting the distribution of $(S_{e}, e \in \path)$, which we summarize in two categories, i) {\it~within-trip (serial)} \textit{dependency}, which refers to the dependency between speed on consecutive edges within the same trip; and ii) {\it~filtration (temporal) dependency}, which refers to the fact that the distribution of speed at an edge, for a specific trip, depends on the arrival time at that edge, and hence on the travel time up to that edge.

To further understand the structural difference between time and serial dependency, let \((U_{e}, e \in \path)\) define a sequence of serially dependent Uniform[0,1] random variables. A trip's distribution of speed over a route \(\path\) is then
\begin{equation}\label{eq:cdf-speed-trip}
  (S_{e}, e\in \path) \stackrel{d}{=} (F^{-1}_{e}(U_{e}, \tau_{e}), e \in \path),
\end{equation}
where times \((\tau_{e}, e\in \path)\) represent the arrival time at edge \(e\). We use the notation \(\tau_{e}\), rather than \(t_{e}\), since the former is a random time. The difference between \eqref{eq:cdf-speed-network} and \eqref{eq:cdf-speed-trip} is that the latter captures extra dependencies associated with vehicle behaviour. {For example, on uncongested highways, drivers can sustain higher-than-average driving speeds for long periods, as there are no traffic events slowing them down. In other words, congestion tends to break within-trip speed correlations.} We refer to this form of dependency as within-trip dependency and associate it with the serial dependencies in \((U_{e}, e\in\path)\). In \eqref{eq:cdf-speed-network} we marginalized out the vehicle behaviour to look at the (unconditional) distribution, {that is,~\eqref{eq:cdf-speed-network} is the distribution of speed of edge $e$ at time $t$, while~\eqref{eq:cdf-speed-trip} is the distribution of speed of a vehicle travelling $e$, conditional on observed speeds of the vehicle on all edges up to $e$.}

Filtration dependency arises from the dependency of speed distribution on time \(\tau_{e}\), as in~\eqref{eq:cdf-speed-trip}. It consequently affects the variability in \(F_{e}(\cdot, t)\) across time. In the real-world, filtration dependency is induced by traffic flow. For example, at night it is safe to assume that all roads are fairly empty, resulting in a time-invariant \(F_{e}\) for that period. On the other hand, filtration dependency is strongest in times of heavy traffic. By accounting for filtration dependency, our developed methods (see Section~\ref{sec:predictive-distribution}) can reduce predictive bias, achieving tighter prediction intervals when compared with
competing methods (Section~\ref{sec:comp-altern-models}).

We pair "\(e \in \path\)" with a random variable to refer to the conditional version, as in \eqref{eq:cdf-speed-trip}, as opposed to "$e \in E$", for the unconditional version, as in \eqref{eq:cdf-speed-network}. We also assume no across-trip dependency, that is, trips are independent from each other. The next section introduces a {form} of serial dependency assumed for \((U_{e}, e \in \path)\), then characterizes travel time as a sampling process over \(G\).

\subsubsection{Travel time as a sampling process} \label{sec:travel-time-sampling}
For generality and empirical purposes, we have not assumed a specific distribution for speed. Nonetheless, various empirical studies have shown that within-trip dependency decreases with distance; see for example~\citet[fig. 5]{woodard2017predicting} and \citet[p. 193]{geroliminis2006real}. 
A sequence of random variables that exhibit such a form of serial dependency, 
with variables far apart being nearly independent, is referred to as a {\it mixing} sequence. Different mixing types have been introduced in the literature. Each mixing type is associated with a separate coefficient assessing its strength~\citep{bradley2005}. The most general, in a sense implied by many other types, is called \(\alpha\){\it -mixing} (strongly mixing), which is defined below.

\begin{definition}[\citep{rosenblatt1956central}] \label{def:alpha-mixing} Let $(X_k, k \in \Z)$ be a sequence of random variables defined on the probability space $(\Omega, \F, P)$. Define the $\sigma$-algebra $\F_a^b$ as
\(\F_a^b = \sigma(X_k, a\leq k \leq b, k \in \Z)\), \(1 \leq a \leq b\leq \infty.\)
For each $n\geq1$ define the measure of dependence
\begin{equation}\label{eq:alpha-mixing}
    \alpha(n)= \sup_{k \geq 1}\sup_{A \in \F_1^k, B \in \F_{k+n}^\infty} |P(A \cap B) - P(A)P(B)|.
\end{equation}
If $\alpha(n) \rightarrow 0$ as $n\rightarrow \infty$, then $(X_k, k \in \Z)$ is said to be {\it $\alpha$-mixing}.
\end{definition}

With this general mixing form of dependency, we assume that the sequence \((U_{e}, e \in \path)\), in~\eqref{eq:cdf-speed-trip}, is \(\alpha\)-mixing. We can equally assume that \((U_{e}, e \in \path)\) is a Markov sequence, which can lead to a simpler and possibly more analytical formulation than the mixing approach. However, empirical evidence of such dependencies is weak~\citep{woodard2017predicting}. Moreover, \(\alpha\)-mixing is a more general form of dependency, in the sense that a Markov sequence is \(\alpha\)-mixing, but the inverse is not true.

Since \((S_{e}, e\in \path)\) are not strictly stationary (meaning that they do not have a time-invariant distribution), we require an extra mixing condition that is slightly stronger than the maximal correlation coefficient (defined as \(\rho(n)\) in \cite{bradley2005}) and implies it, which is related to mixing of interlaced sets.
\begin{definition} \label{def:rho-mixing} Following 
Definition 
\ref{def:alpha-mixing}, let \(\F_{A} =\sigma(X_{i}, i \in A)\) for any non-empty set \(A \subset \Z\). Define the dependency measure
\begin{equation}
  \label{eq:rho-mixing}
  \trho (n)= \sup_{A, B}\sup_{f,g} |\Corr(f,g)|, \quad  f \in \L^{2}(\F_{A}),\; g \in \L^{2}(\F_{B}), \min_{i\in A,j\in B }|i-j| \geq n,
\end{equation}
where \(\L^{2}(\F_{A})\) denotes the space of square-integrable \(\F_{A}\) random variables, and the supremum is over all disjoint non-empty sets \(A, B \in \Z\).
\end{definition}

Generally, we have that \(0\leq \trho(n) \leq 1\). Hence, the objective of Definition \ref{def:rho-mixing} is to define a dependency measure that ensures that there are no fully correlated disjoint sets of random variables. In real-world networks, speeds on disjoint edges are correlated, but not fully correlated, making this condition plausible. From mixing Definitions \ref{def:alpha-mixing} and \ref{def:rho-mixing}, we let \((S_{e}, e \in \path)\) be sequential samples from edges in \(\path\) over a transportation network \(G\), such that, for a given route, \(\path = \langle e_{1}, e_{2}, \dots, e_{n}\rangle\),
the sampling occurs at the random arrival times \( \tau_{e_{1}} < \tau_{e_{2}}< \dots <\tau_{e_{n}}\), defined as
\begin{equation}\label{eq:random-arrival-time}
\tau_{e}= \min\{t >0 : \T_{\langle.., e \rangle} \leq t\},
\end{equation}
or, equivalently, through the recursive relation defined later in~\eqref{eq:recursive-map}. Here \(\langle.., e \rangle\) is the route up to edge \(e\). We define travel time as a random variable as follows. 

\begin{definition}[Travel time random variable] \label{def:travel-time} 
  For a transportation network \(G\), an arbitrary route \(\path\) and a start time \(t_{0}\), let \((U_{e}, e \in \path)\) be an \(\alpha\)-mixing sequence of Uniform[0,1] random variables, such that \(\sum_{n>0}n^{-1}\alpha(n)<\infty\) and \(\lim_{n \to \infty} \trho(n) <1\). Let \(U_{\langle.., e' \rangle} = (U_{e}, e \in \langle.., e' \rangle )\). Travel time, as a random variable, is constructed as
\begin{equation}
  \label{eq:trave-time-decomposition-time-series}
  \T_{\path} = \sum_{e \in \path} d_{e}m_{e}(\tau) + \sum_{e \in \path} d_{e}\epsilon_{e}(\tau),
\end{equation}
where \(m_{e}(\tau) = \E[S_{e}(\tau) \mid U_{\langle.., e \rangle}]\), \(\epsilon_{e}(\tau) = S_{e}(\tau) -m_{e}(\tau)\), \(\tau_{e}\) as in \eqref{eq:random-arrival-time}.
\end{definition}

In~\eqref{eq:trave-time-decomposition-time-series}, we removed the sub-index $_e$ of \(\tau\) since it is already implied by the subscript of the functional. From Definition \ref{assump:seasonality-of-speed-on-a-road}, the residual \((\epsilon_{e} (\tau), e \in \path) \) are not identically distributed. They are also dependent through \(U_{\langle.., e' \rangle}\). This brings us to the study of long-term behaviour of travel time.
  
\subsection{Asymptotic properties of travel time}\label{sec:asymptotic-properties}
\subsubsection{Asymptotic distribution} \label{sec:asymptotic-distribution}
Estimating the long-term behaviour of the travel time, of Definition~\ref{def:travel-time}, requires proper treatment of filtration dependency. Given \(U_{\langle \dots, e\rangle}\), the expected value of \(\T_{\path}\), $\mu_{\path}(\tau)$, is constructed by conditioning on its own stopping-times \((\tau_{e}, e\in \path)\), as
\begin{equation}
  \label{eq:online-mean}
  \mu_{\path}(\tau) = \sum_{e \in \path}d_{e}\E[S_{e}(\tau) \mid U_{\langle \dots, e\rangle}] = \sum_{e \in \path} d_{e}m_{e}(\tau) +\sum_{e \in \path}d_{e}\E[\epsilon_{e}(\tau)\mid U_{\langle \dots, e\rangle}].
\end{equation}

The exact value of \(m_{e}(\tau)\) in~\eqref{eq:online-mean} is only known at time \(\tau_{e}\). Hence, \(\mu_{\path}(\tau)\) is updated at each edge \(e \in \path\), in an online way. Similarly, the variance of \(\T_{\path}\) is
\begin{equation}
  \label{eq:online-variance}
  \sigma^{2}_{\path}(\tau)= \sum_{e \in \path}d_{e}^{2}\sigma^{2}_{e}(\tau) + \sum_{e, e' \in \path}d_{e}d_{e'}\Var(\epsilon_{e}(\tau), \epsilon_{e'}(\tau)\mid U_{\langle \dots, e\rangle} ),
\end{equation}
where \(\sigma^{2}_{e}(\tau) = \Var(\epsilon^{2}_{e}(\tau)\mid U_{\langle \dots, e\rangle})\), is the edge-level variance. Our first result states that the average travel time for arbitrary routes on the network converges asymptotically to a constant that is independent of initial conditions (i.e.~start time) and route.

\begin{lemma} \label{thm:slln-uniform} With Definition \ref{assump:seasonality-of-speed-on-a-road}, let \(\path\) be a random walk on \(G\). Let \(\T_{\path}\) be in Definition~\ref{def:travel-time}, then \(n^{-1} \T_{\path}\to \mu\) almost surely~(a.s.) as \(n\to \infty\), where \(\mu\) is the invariant expected {travel time for an arbitrary edge}, defined as \(\mu = \sum_{e \in E}\spi_{e}\mu_{e} \), with \(\mu_{e} = d_{e}\E[S_{e}]= d_{e}\int m_{e}(t)dt\), the unconditional average travel time over \(e\), and \(\spi_{e} =n_{e}/n\) as \(n\to \infty\), the stationary probability of travelling \(e\).
\end{lemma}

Because travel time is an empirical process, and in the view of ride-share providers, where vehicles are continually and randomly assigned rides to arbitrary locations on the network, we built on the fact that \(\path\) is a random walk on \(G\) in Lemma~\ref{thm:slln-uniform}. Many deterministic systems are essentially random walks in the limit. For example, taking a right turn on every node on a \(d\)-degree graph, where every node is with \(d\)-edges, is a random walk \citep{aldous}. If \(\path\) is cyclical, the result of Lemma~\ref{thm:slln-uniform} still holds, since the subgraph constructed from the cycle is still a graph, and \(\mu\) \mo{depends on this subgraph}.

The main proof idea in Lemma~\ref{thm:slln-uniform} comes from the dynamical system's literature. The edge-arrival times \((\tau_{e}, e\in \path)\) defined in~\eqref{eq:random-arrival-time}, can equally be represented by the recursive map
\begin{equation} \label{eq:recursive-map}
\tau_{e} = \tau_{e'} + d_{e'}S_{e'}(\tau_{e'}),
\end{equation}
where \(\langle e', e \rangle\) is a consecutive sequence of edges in \(\path\). The map~\eqref{eq:recursive-map} is known as the rotation mapping and is studied under deterministic settings, that is when \(d_{e'}S_{e'}(\tau_{e})\) in~\eqref{eq:recursive-map} is replaced by a fixed constant~\citep[Prop 2.16]{einsiedler2013ergodic}. To show the benefit of the representation in~\eqref{eq:recursive-map}, consider a sufficiently smooth and periodic function \(f(x)\), which can always be defined over its period, say \([0, 1]\) for simplicity, such that for any \(x \in \R\), \(f(x) = f(x \pmod 1)\). Then, for any sequence \((x_{0}, x_{1}, x_{2}, \dots)\) defined by a rotation map \(x_{k} = T^{k-1}(x) = x_{k-1} + \alpha \pmod 1\), \(k =1, 2 \dots\), for some constant \(\alpha\), if the map \(T(x)\) is ergodic, then
\begin{equation}
\label{eq:average-ergodic}
\frac{1}{n}\sum_{k=1}^{n}f(T^{k}(x)) \rightarrow\int_{0}^{1}f(x)dx,\quad \text{a.s.~as }  n \to \infty.
\end{equation}

Ergodicity implies that the map \(T(x)\) forgets its initial starting point (or long-term memory) as the number of steps \((k)\) increases. In this sense, the right-hand side of~\eqref{eq:average-ergodic} does not depend on the starting value \(x_{0}\), and all averages initiated from arbitrary starting values would converge to the same right-hand side constant, under the same rotation map. Letting \(f\) be \(m_{e}\), for each edge \(e\), we show that the mapping \eqref{eq:recursive-map} is ergodic, even though it depends on previous arrival times. Hence, \(n_{e}^{-1}\sum_{i=1}^{n_{e}}m_{e}(\tau_{i})\) converges to an edge-specific constant \(\mu_{e}\) representing the unconditional expected travel time of \(e\), where \(\tau_{i}\) is the \(i\)th arrival time to \(e\). See Supplementary Material (SM) Section \ref{app:proof-lemma-slln-stationary} for a more detailed proof.

Our next result shows that the rotation mapping in~\eqref{eq:recursive-map} is also mixing, in the sense that a \(\sqrt{n}\)-normalization of \(\T_{\path}\) causes the deviations around \(\mu\) to behave like deviations from a normal distribution. Deterministic rotations can be ergodic, but they are not mixing.

Motivated by the work on mixing random variables in \cite{peligrad1996asymptotic}, we establish a central limit theorem (CLT) for travel time, with proof in SM Section \ref{app:proof-clt-offline-online}.
\begin{theorem}[CLT for travel time] \label{thm:clt-traveltime-offline} Following the settings of Lemma \eqref{thm:slln-uniform}, let \(\mu\) be the invariant expected travel time. Then, \(n^{-1} \sigma^{2}_{\path}(\tau) \to \sigma^{2}\), a constant. If \(\sigma^{2} \neq 0\), then
    \begin{equation}
      \label{eq:clt-offline}
      n^{-1/2}(\T_{\path}- n\mu) \xrightarrow d N(0,\sigma^{2}), \quad \text{as } n \to \infty.
    \end{equation}
    Both \(\mu\) and \(\sigma^{2}\) are independent of initial conditions and \(\path\).
\end{theorem}

The \(\xrightarrow d\) refers to convergence in distribution. Regardless of start time and route, Theorem~\ref{thm:clt-traveltime-offline} states that the longer the trip is, the closer the average travel time is to a single universal constant \(\mu\) (Appendix Fig~\ref{fig:hatmu-different-sampling-methods} illustrates this phenomenon in Quebec City Data), with deviations based on a universal constant \(\sigma\). The condition that \(\sigma^{2}\not = 0\) is not as stringent in real-world transportation networks, since speed limits vary across edges. 

By estimating \((\mu, \sigma)\), which is the focus of the next section, Theorem \ref{thm:clt-traveltime-offline} enables us to quantify travel uncertainty, such as quantifying the probability of the event \(\{\T_{\path} < t\}\), \(t>0\).
\subsubsection{Estimation of $(\mu, \sigma)$}\label{sec:estimation-mu-sigma}
By the cyclostationarity of \(G\), the expected value of the average travel time over an arbitrary route of length \(n\) is \(\mu\), as \(\E[n^{-1}\T_{\path} \mid n]=\mu\), for all \(n \in \Z\). This expectation is with respect to the stationary distribution \((\spi_{e}, e \in E)\). Transportation data is composed of arbitrary trips, with differing routes on the network, therefore, we treat \(n\) as a random variable. By the law of total variance, the unconditional variance is
\begin{equation}\label{eq:uncoditional-variance-traveltime}
  \begin{aligned}
    \Var(n^{-1}\T_{\path}) &= \E[ \Var(n^{-1}\T_{\path} \mid n)] + \Var(\E[n^{-1}\T_{\path}\mid n])  \\ &=\E[n^{-1}(\sigma^{2} + O(n))] + \Var(\mu) = \sigma^{2}\E[n^{-1}].
  \end{aligned}
\end{equation}

With a slight abuse of notation, \(O(n)\) represents the residual as a random variable of the average variance of travel time, as \(n^{-1}\Var(\T_{\path}) -  \sigma^{2} =  O(n)\). The expectation is with respect to distance as \(\E[n^{-1}O(n)] = \sum_{n_{0}}\P(n = n_{0})n_{0}^{-1}\E[O(n_{0})]=0\), since the latter is an expectation with respect to time.

Hence, given a representative independent sample of \(m\) trips \(\T_{\path}^{(j)},\; j=1, \dots, m\), with \(n_{j}\) edges each, an unbiased estimator of \(\mu\) is
\begin{equation}
  \label{eq:sample-mean}
  \hatmu = \frac{1}{m}\sum_{j=1}^{m}\frac{\T_{\path}^{{(j)}}}{n_{j}}.
\end{equation}

{Here, $\path$ in $\T_{\path}^{(j)}$ is trip dependent, i.e.~$\path^{(j)}$, however we suppressed the superscript to simplify the notations}. By conditioning on \(n_{j}\), with the laws of total expectation and variance, we have that \(\E[\hatmu] = m^{-1}\sum_{j=1}^{m}\E\big [\E [n_{j}^{-1}\T_{\path}^{{(j)}} \mid n_{j}=n  ]\big ] = \mu\), and
\begin{equation} \label{eq:variance-sample-mean}
  \begin{aligned}
    \Var(\hatmu) &= \E[\Var(\hatmu \mid n)] + \Var(\E[\hatmu \mid n]) \\ &=  \frac{1}{m^{2}}\sum_{j=1}^{m}\E\bigg [\frac{1}{n} \big [\sigma^{2} + O(n)] \bigg ]+  \Var ( \mu) = \frac{\sigma^{2}}{m}\E[n^{-1}].
  \end{aligned}
\end{equation}

 For a fixed route length, such as \(n_{j}=n\) for all \(j=1, \dots ,m\), \(\Var(\hatmu) = \{mn\}^{-1}\sigma^{2}\). If \(n=1\) we retrieve the classical sample variance \(m^{{-1}}\sigma^{2}\). Applying the classical results of the central limit theorem of the sample mean, we have
\begin{equation}
  \label{eq:convergence-sample-mean}
    \sqrt{m}(\hatmu - \mu) \xrightarrow d N \big (0,\sigma^2\E[n^{-1}] \big ), \quad \text{as } m \to \infty.
\end{equation}

From \eqref{eq:uncoditional-variance-traveltime} and \(\E[n^{-1}\T_{\path}\mid n]=\mu\), a consistent and unbiased estimator of the unconditional variance \(\sigma^{2}\E[n^{-1}]\) is the sample variance, as 
\begin{equation}
  \label{eq:mse-sample-mean}
  \sighatmu = \frac{1}{m-1}\sum_{j=1}^{m}(n_{j}^{{-1}}\T_{\path}^{(j)} - \hatmu)^{2}. 
\end{equation}

Since \(n_{j}^{-1}T_{\path}^{(j)}\) are independent and identically normally distributed samples over \(G\), then \(\sighatmu\) are distributed as a chi-square with \(m-1\) degrees of freedom \citep[Thm. 5.3.1]{casella2002statistical}. Moreover, \(\sighatmu^{-1/2}(\mu -\hatmu) \stackrel{d}{=} T^{(m-1)}\), where \(T^{(m)}\) is a student-t distribution with \(m\) degrees of freedom \citep[Sec. 5.3.2]{casella2002statistical}. The variance\footnote{By assuming weak stationarity of the variance, following the argument of \citet[page 99]{herrndorf1983invariance}, the variance can be represented as  \(\sigma^{2}(n) = nh(n)\), where \(n\) is the length of the route, and \(h(n)\) is a slow varying function. By Karamata representation theorem for slow varying function, $h$ can be represented as \(h(n) = \exp\big (f(n) + \int_{0}^{n}t^{-1}g(t)dt \big)\), for two bounded measurable functions \(f\) and \(g\), where \(f(n)\) converges to a constant and \(g(n)\) to zero, as \(n \to \infty\). This constitutes an alternative approach to modelling the variance.} \(\sigma^{2}\) in Theorem~\ref{thm:clt-traveltime-offline} represents the limit of the conditional variance \(n^{-1}\Var(\T_{\path}\mid n)\), while \(\Var(n^{-1}\T_{\path}) = m\Var(\hatmu)\) is the total variance that treats \(n\) as a random quantity. Let \(\hatn = m^{-1}\sum_{j=1}^{m}n_{j}^{-1}\), from \eqref{eq:mse-sample-mean}, a profile estimator of \(\sigma^{2}\) is
\begin{equation}
  \label{eq:profile-estimator}
  \sigmaprofile^{2} =\frac{\sighatmu}{\hatn}.
\end{equation}
\subsubsection{Confidence intervals}\label{sec:conf-bounds}
The normality result in Theorem~\ref{thm:clt-traveltime-offline} and the mean and variance estimators of the previous section allow confidence intervals for the average travel time \(\mu\) to be constructed easily. For a large sample of \(m\) trips, from \eqref{eq:convergence-sample-mean}, a \((1-\qparam)\)100\%, \(\qparam \in (0,1)\), confidence interval for $\mu$ is
\begin{equation} \label{eq:mean-confidence-intervals}
\mu \in \Bigg [\hatmu - T^{(m-1)}_{\qparam/2}\sqrt{\frac{\widehat\Var(n^{-1}\T_{\path})}{m}} ,\quad \hatmu + T^{(m-1)}_{1-\qparam/2} \sqrt{\frac{\widehat\Var(n^{-1}\T_{\path})}{m}} \,\Bigg ],
\end{equation}
where \(\hatmu\) as in \eqref{eq:sample-mean}, \(\widehat\Var(n^{-1}\T_{\path})\) as in \eqref{eq:mse-sample-mean}, and \(T^{(m)}_{\qparam}\) is the \(\qparam\)-quantile of a student-t distribution with \(m\) degrees of freedom. 

\subsubsection{Population prediction intervals}\label{sec:point-wise-asympt}
From \eqref{eq:clt-offline}, we know that \(\Var(\Tnew_{\path})= n\sigma^{2}\). When the mean and variance are known, the \((1-\qparam)\)100\% intervals of \(N(n\mu, n\sigma^{2})\) distribution can be used as a prediction interval.  When the mean is unknown and the predictor of \(\Tnew_{\path}\) is \(n\hatmu\), a prediction interval must take into account predictor uncertainty \citep{geisser2017predictive}. The route-length conditional variance is \(\Var(\Tnew_{\path} - n\hatmu \mid n) = n\sigma^{2}(1+ m^{-1}) \). Using the profile estimator \(\sigmaprofile^{2}\) of~\eqref{eq:profile-estimator}, we have \(\widehat  {\Var}(\Tnew_{\path} - n\hatmu \mid n) = n\sigmaprofile^{2} (1 + m^{-1}) \). By accounting for predictive uncertainty, a point-wise asymptotic prediction interval is of the form
\begin{equation} \label{eq:prediction-interval}
\Tnew_{\path} \in \Bigg [n\hatmu - z_{\qparam/2}\sqrt{n \sigmaprofile^{2} \big ( 1+\frac{1}{m} \big )} ,\quad n\hatmu + z_{1-\qparam/2}\sqrt{n \sigmaprofile^{2} \big (1+\frac{1}{m} \big )} \,\Bigg ],
\end{equation}
where \(z_{\qparam} = \inf\{ x \in \R : 1- \Phi(x)> \qparam\}\), and \(\Phi(x)\) is the cumulative distribution function of a standard normal random variable.

By conditioning the variance estimate on \(n\),~\eqref{eq:prediction-interval} is a population interval, in the sense that it will cover with \((1-\qparam)\%100\) level of significance any arbitrary route of \(n\) edges from the population of routes of that length. This follows from the fact that both \(\hatmu\) and \(\sigmaprofile\) are estimated from pooled trips, in~\eqref{eq:sample-mean} and~\eqref{eq:mse-sample-mean}, respectively.

To use~\eqref{eq:prediction-interval} for route-specific (and possibly time) prediction intervals, one would need a sample of \(m\) independent trips of the same route (and start-time) to calculate the parameters \(\hatmu\) and \(\sigmaprofile^{2}\) used in \eqref{eq:prediction-interval}. The next section illustrates another approach, by constructing trip-specific mean and variance sequences that centre and normalize travel time to a Gaussian-based predictive distribution, achieving tighter prediction bounds.

\subsection{Trip-specific predictive distribution}\label{sec:predictive-distribution}
Most applications are interested in bounding travel time by constructing predictive intervals. Different types of intervals are suitable for different objectives. Population estimators, as in the universal parameters \((\mu, \sigma)\) of Section~\ref{sec:estimation-mu-sigma}, provide asymptotic bounds in \eqref{eq:prediction-interval} that are wide on the short-term and converge to zero in the long-term when considering \(n^{-1}\Tnew_{\path}\). This section provides predictive interval sequences that are trip-specific, tighter on the short-term but whose length does not converge to zero as \(n\to\infty\).

Given a route $\path$, and start time $t_0$, using the recursive approach in~\eqref{eq:recursive-map}, calculate the deterministic arrival times \(\Tstar = (\Tstar_{e}, e\in \path)\) at every edge in the route, as
\begin{equation}
  \label{eq:deterministic-times}
  \Tstar_{e} = \Tstar_{e'} + d_{e'} m_{e'}(\Tstar_{e'}), \qquad \langle e', e \rangle \in \path.
\end{equation}

We use \(t\) rather than \(\tau\) to refer to the deterministic nature of \(\Tstar\). For example, if $e''$ is the first edge in $\path$, then set $\Tstar_{e''} = t_0$, and recursively generate all \((\Tstar_{e}, e\in \path)\). Hence, for any given route $\path$ and start time $t_0$, we calculate the expected travel time \(\mu_{\path}(t_0)\) as
\begin{equation}
  \label{eq:mu-recursive}
   \mu_\path (t_0) = \sum_{e \in \path}d_e m_e(\Tstar_e).
\end{equation}

Constructing a covariance sum that is similar to~\eqref{eq:online-variance}, using \((\Tstar_{e}, e\in \path)\),  requires the estimation of \(2^{-1}n(n+1)\) terms: \(n\) edge\(\times\)time specific variances and \(2^{-1}n(n-1)\) pairwise correlation coefficients. This is a daunting task. A reduced covariance sum, which is profiled at a single correlation value and only requires \(n+1\) parameters, can be used instead, such as
\begin{equation}
  \label{eq:sigma-reduced}
  \sigma^{2}_{\path}(t_0)= \sum_{e \in \path}d_{e}^{2}\sigma^{2}_{e}(\Tstar_e) + 2\xiprof \sum_{\langle e, e'\rangle \in \path}d_{e}d_{e'}\sigma_{e}(\Tstar_e) \sigma_{e'}(\Tstar_{e'}),
\end{equation}
where \(\xiprof\) is a proxy to the average lag-one auto-correlation over \(G\), and \(\sigma_{e}(\Tstar_e)\) is the variance at the deterministic times in \eqref{eq:deterministic-times}. We use the subscript \(_{\text{prof}}\), since the variance in \eqref{eq:sigma-reduced} is profiled at the value \(\xiprof\), rather than the true values of each pair-edge correlation.

From \eqref{eq:mu-recursive} and \eqref{eq:sigma-reduced}, we define the asymptotic predictive distribution of travel time. It is predictive in the sense that it predicts the distribution of a trip apriori, and thus contains an added noise source resulting in an extra variance. 
\begin{theorem}[Predictive distribution of $\T_{\path}$ ]\label{thm:predictive-traveltime}
Following the settings of Lemma \ref{thm:slln-uniform}, for an arbitrary start time $t_0>0$, let \(\mu_{\path}(t_0)\) be as in \eqref{eq:mu-recursive} and \(\sigma_{\path}(t_0)\) as in \eqref{eq:sigma-reduced}, then
  \begin{equation} \label{eq:predictive-clt}
    \sigma_{\path}^{{-1}}(t_0)(\T_{\path} - \mu_{\path}(t_0)) \xrightarrow d \sqrt{\eta}N(0,1 + \tilde\sigma^{2}), \quad  \text{as } n\to\infty,
  \end{equation}
  where \(\eta\) is a strictly positive constant representing the ratio of \(n\sigma^{2}\) to \(\sigma^{2}_{\path}(t_0)\), where $\sigma^2$ is the population variance in~\eqref{eq:clt-offline}. Moreover, $\eta$ and $\tilde \sigma^2$ are independent from start time $t_0$ and $\path$. The constant $\tilde \sigma^2$ is defined as
  \[\tilde \sigma^{2} = \E[d_e^2\Var(m_{e}(t) \mid e)]=\sum_{e \in E} \spi_{e} d_e^2 \bigg [\int m_{e}^{2}(t)dt - \bigg (\int m_{e}(t)dt\bigg )^{2} \bigg ].\]
\end{theorem}

The main idea for the proof of Theorem~\ref{thm:predictive-traveltime} is to decompose the left hand of~\eqref{eq:predictive-clt} as
\[\frac{\T_{\path} - \mu_{\path}(t_0)}{\sigma_{\path}(t_0)}
    = \frac{\sqrt{n}\sigma}{\sigma_{\path}(\Tstar)} \frac{\T_{\path} - \mu_{\path}(\tau)}{\sqrt{n} \sigma} -  \frac{\sqrt{n}\sigma}{\sigma_{\path}(t_0)}\frac{\mu_{\path}(t_0) - \mu_{\path}(\tau)}{\sqrt{n}\sigma}= I  - II.\]

  By Slutsky's theorem and Theorem \ref{thm:clt-traveltime-offline} we have \(I \xrightarrow d  \sqrt{\eta} N(0,1)\). We further show that the mapping~\eqref{eq:deterministic-times} is ergodic and mixing (as in Lemma~\ref{thm:slln-uniform}), such that \(II \xrightarrow d \sqrt{\eta}N(0, \tilde\sigma^{2})\). Since we assumed that different trips are independent (cross-trip dependency), then \(I\independent II\), establishing the desired results. See SM Section~\ref{app:proof:predictive-distribution} for a detailed proof.

Estimates of \(\big (\{m_e , \sigma_{e}\},  e\in E )\), as well as $\{\xiprof, \eta, \tilde \sigma^2,\mu_\path(t_0), \sigma_\path(t_0)\}$ are required to construct a prediction interval for a new trip \(\Tnew_\path\).  To carry this, let \(\big (\{\hatm_e , \hatsigma_{e}\},  e\in E \big )\) be the sample means and variances of speed for every edge in $E$. Time bins can be used, for example, every edge can have 168 mean and variance estimates for every week hour or 24 for day hours. While the proposed methods do not depend on the time bin choice, the variability of traffic estimates trickles into travel time uncertainty. 

Let \(\T_{\path}^{(j)},\; j=1, \dots, m\) be a sample of $m$ independent trips, each with route \(\path^{(j)}\) of $n_{j}$ length. Let \((s^{(j)}_e, e \in \path^{(j)})\) be the sequence of observed speeds of trip $j$ in order of travel, where $(\{\hatm_e^{(j)}, \hatsd_{e}^{(j)}\}, e\in \path^{(j)})$ are the estimates of the average and standard deviation of speed at every edge on the path $\path^{(j)}$ when the corresponding speed was observed. A population estimator of \(\xiprof\), pooling from multiple trips, is
\begin{equation}
  \label{eq:xi-estimate}
  \hatxi  = \frac{1}{m} \sum_{j=1}^{m}  \frac{1}{n_{j}}\sum_{\langle e, e' \rangle \in \path^{(j)}}\frac{ \big ( s^{(j)}_{e} - \hatm^{(j)}_{e} \big )  \big ( s^{(j)}_{e'} - \hatm_{e'}^{(j)}\big )}{\hatsd_{e}^{(j)} \hatsd^{(j)}_{e'}}.
\end{equation}

Even though $\eta$ and $\tilde \sigma^{2}$ in~\eqref{eq:predictive-clt} are well-defined quantities, their estimators can be hard to compute. A classical sample variance estimator can be used for the conditional variance of \(\Var(m_{e}(t)\, \mid e)\) if large amounts of data per edge at time \(t\) are available. Otherwise, one requires smoothing or time binning. Therefore, we propose a population estimator of the total variance \(\nu^{2}  = \eta(1+\tilde \sigma^{2})\) based on the sample variance of the residual of the $m$ trips, as
\begin{equation}
  \label{eq:sample-variance-residual}
  \hatnu^{2} = \frac{1}{m-1} \sum_{j=1}^{m}(\varepsilon^{(j)} - \bar \varepsilon)^{2}, 
\end{equation}
where \(\varepsilon^{(j)} = \{\hatsd^{(j)}_{\path}(t_0^{(j)})\}^{-1}(\T_{\path}^{(j)} - \hatmu_{\path}^{(j)}(t_0^{(j)}))\), \(\bar \varepsilon = m^{-1}\sum_{j=1}^{m}\varepsilon^{(j)}\), and \(\hatmu_{\path}^{(j)}(t_0^{(j)}))\), $\hatsd^{(j)}_{\path}(t_0^{(j)})$ are the $\T_\path^{(j)}$ specific mean and variance estimates of~\eqref{eq:mu-recursive} and~\eqref{eq:sigma-reduced}, respectively, at the trip's start time $t_0^{(j)}$. For simplicity, we illustrate how to estimate the trip-specific mean and variance for a generic trip, that is, for an arbitrary route and start time $\{\path, t_0\}$. We first compute \((\Tstar_{e}, e \in \path)\) as in~\eqref{eq:deterministic-times}. An estimator of $\mu_\path(t_0)$ is
\begin{equation}
  \label{eq:estimator-mu-recursive}
  \hatmu_{\path}(t_0) = \sum_{e \in \path} d_e\hatm_{e}(\Tstar_e).
\end{equation}

A profile estimator of \(\sigma_{\path}(t_0)\), profiled at \(\hatxi\), is
\begin{equation}
\label{eq:estimator-sigma-reduced}
  \hatsigma_{\path}(t_0)= \sum_{e \in \path}d_{e}^{2}\hatsigma_{e}(\Tstar_e) + 2\hatxi \sum_{\langle e, e'\rangle \in \path}d_{e}d_{e'}\hatsd_{e}(\Tstar_e) \hatsd_{e'}(\Tstar_{e'}).
\end{equation}

From Theorem~\ref{thm:predictive-traveltime},~the estimators \eqref{eq:xi-estimate} and \eqref{eq:sample-variance-residual}, a \(100\times (1-\qparam)\)\% prediction interval for a new trip $\Tnew_{\path}$ starting at $t_0$ for a given route $\path$, is
\begin{equation} \label{eq:prediction-sequence}
\Tnew_{\path} \in \Bigg [\hatmu_{\path}(t_0) - z_{\qparam/2}\sqrt{\hatnu^{2} \hatsd^{2}_{\path}(t_0)},\quad \hatmu_{\path}(t_0) + z_{1-\qparam/2}\sqrt{\hatnu^{2} \hatsd^{2}_{\path}(t_0)} \,\Bigg ],
\end{equation}

See Appendix Algorithm~\ref{algo:trip-specific} for a summary of this estimation procedure. The intervals~\eqref{eq:prediction-sequence} are not the tightest trip-specific intervals, since they include population estimates for  both \(\hatnu\) and \(\hatxi\). Nonetheless, they are tighter for short trips than~\eqref{eq:prediction-interval}, since \(\hatnu\) and \(\hatxi\) pool specific quantities of the variance and not the whole variance as in~\eqref{eq:mse-sample-mean} and~\eqref{eq:profile-estimator}. It is possible to add higher-order covariance terms to the sum in \eqref{eq:sigma-reduced}, and correspondingly to~\eqref{eq:estimator-sigma-reduced}, such as the second-lag auto-correlation coefficient of the residual. This will reduce the variability resulting from pooling in \(\hatnu\), and consequently lead to tighter prediction intervals than~\eqref{eq:prediction-sequence}. Nonetheless, this would require the estimation of additional auto-correlation parameters; thus, its utility is application specific (see Fig~\ref{fig:comparision-of-prediction-intervals}).

\section{Data analysis} \label{sec:case-study} 
\subsection{Outline}
We start by exploring the data and parameter estimates for the different traffic-bins in sections~\ref{sec:traff-bin-estim} and \ref{sec:parameter-estimation}. Section~\ref{sec:pool-based-prediction-intervals} evaluates and compares the population prediction intervals to the trip-specific prediction sequences. We end, in Section~\ref{sec:comp-altern-models}, by comparing the out-of-sample performance of our method against previously proposed methods for travel time estimation that also quantify travel uncertainty. 

\subsection{Traffic-bin estimators}\label{sec:traff-bin-estim}
{
The estimates \((\hatm_{e}, \hatsigma_{e}, e \in E)\) are calculated using the training set. A unit is defined as edge\(\times\)traffic-bin, using the traffic-bins introduced in Section \ref{subsec:split}. Since all our notations use the path-conditioning \((e \in \path)\), the sample means of speed per unit are calculated for every exit of an edge. For example, if edge $e$ has two exits $\{e_1, e_2\}$, then $e$ has three estimates per unit, one for each pair \(\langle e, e_{i} \rangle\), $i\in \{1, 2\}$, by considering observations over $e$ conditional that the trip took exit $e_i$, and one unit-level estimate that ignores the exit. For a trip travelling \(\langle e_{1}, e_{2}, e_{3} \rangle\), \(\hatm_{e_{1}}\), the assigned traffic estimates would correspond to \(\langle e_1, e_2 \rangle\), \(\langle e_2, e_3 \rangle\) and the unit-level estimate \(\langle e_3 \rangle\) (respecting traffic-bins). For an estimate to be used in practice, we require at least 10 observations. Approximately 90\% of estimates have less than 10 observations. We impute those estimates in the following order. First, a) impute with the unit-level estimate (ignoring exits) if the path-conditional estimate has insufficient data; otherwise b) impute with the traffic-bin estimate when a) has insufficient data. The latter is an estimate that uses all data within a traffic-bin (ignoring the edge). Even though this imputation procedure is crude, the results are promising.}

\subsection{Parameter estimation}\label{sec:parameter-estimation}
To understand how traffic patterns reflect on our methods, Table~\ref{tb:parameter-estimates} reports parameter estimates of population prediction intervals~\eqref{eq:prediction-interval}, and the trip-specific sequences~\eqref{eq:prediction-sequence}, for the sampling at random approach (from the entire data) and the AM-rush and Non-rush strata (Sec.~\ref{subsec:split}). We focus on those two strata (AM/Non) because of traffic similarities between the AM and PM-rushes.

Parameter estimates (\(\hatmu, \sighatmu, \hatn, \sigmaprofile\)) are calculated from 1,000 trips sampled at-random from the training data according to each stratum. We also report, in parentheses, the 95\% confidence intervals for \(\hatmu\) calculated according to \eqref{eq:mean-confidence-intervals}. We found that sampling more than 1,000 trips did not lead to significantly different estimates than those in Table~\ref{tb:parameter-estimates}.
\begin{table}[!htp]
  \centering
    \caption{Parameter estimation under different sampling methods}
    \label{tb:parameter-estimates}
    {\footnotesize 
  \begin{tabular}{l lllll}
                      & \multicolumn{3}{c}{Sampling method}                                \\
                      & At random         & \multicolumn{2}{c}{Stratified by traffic-bins} \\[5pt]
                     
                      &                   & {AM-rush} & {Non-rush}                         \\
    \(\hatmu\)        & 16.70 (16.4,17.1) & 17.30     & 13.90                              \\   
    \(\sighatmu\)     & 33.50             & 37.90     & 22.70                              \\   
    \(\hatn\)         & 0.02              & 0.02      & 0.02                               \\        
    \(\sigmaprofile\) & 41.80             & 44.40     & 32.40                              \\
    \midrule
    \(\hatxi\)        & 0.31              & 0.29      & 0.33                               \\
    \(\hatnu\)        & 1.28              & 1.46      & 1.09
  \end{tabular}
  }
\end{table}

For the trip-specific parameters of Section~\ref{sec:predictive-distribution}, we require an estimate of the edge-specific mean and variance pair \(\{\hatm_{e}, \hatsigma_{e}\}\) for each traffic-bin, and hence we use the whole training data  to estimate those parameters. For $\hatxi$ and $\hatnu$, we estimate them for the AM-rush and Non-rush strata from 4,000 trips sampled randomly from each stratum, respectively (reported in Table~\ref{tb:parameter-estimates}). Another 4,000 randomly selected trips from the whole training data are used to estimate $\hatxi$ and $\hatnu$ to resemble a sampling-at-random approach. Trips overlapping more than one traffic-bin were removed from parameter estimation in Table~\ref{tb:parameter-estimates}. 

Traffic patterns affect both the average travel time, on an arbitrary edge and its variability. The average travel time for an arbitrary edge is approximately 17.3 seconds in the AM-rush stratum, with a standard deviation of \(\sigmaprofile = 44.40\). Travel is faster and more certain in the Non-rush stratum than in the AM-rush, with an average of about 13.9 seconds and a standard deviation of 32.40 seconds. Traffic also affects the lag-one auto-correlation \(\hatxi\) and the residual variability. In non-rush hours, travel time is slightly more correlated (0.33) with less variability (1.09) than the AM-rush hours {(0.29 and 1.46, respectively).} {This is expected in real-world road networks: higher traffic can result in more road events than usual that break the correlation within the speeds of a trip and increase the variability of travel time.} Moreover, the average of \(\hatxi\) across all trips in the training set is 0.3 (see SM Figure \ref{app:fig:lag-1-autocorrelation}). {If $\hatmu,\sigmaprofile $ is to be plotted against the number of vehicles in the network (or any proxy of $\Pi)$, we believe the shape would be similar to the network-wide fundamental diagram of~\cite[Fig 7]{geroliminis2008existence}. However, we lack sufficient data to do so.}

Our methods build on the assumption of ergodicity of the system, meaning that the average travel time (\(\hatmu\)) can be estimated by an average of travel time for a single very long trip (time average), and equally by averaging the average of multiple trips (space average) as in~\eqref{eq:sample-mean}. We illustrate this phenomenon empirically in Appendix Figure~\ref{fig:hatmu-different-sampling-methods} showing that space and time averages are almost {equal} for all lengths \(n\), well within the 95\% confidence intervals in Table~\ref{tb:parameter-estimates}. 

\subsection{Comparison of population and trip-specific intervals}\label{sec:pool-based-prediction-intervals}
To illustrate the advantage of accounting for filtration dependency through the rotation map~\eqref{eq:recursive-map}, we compare our trip-specific sequences~\eqref{eq:prediction-sequence}, first graphically and then numerically, to the asymptotic intervals~(\ref{eq:prediction-interval}). Our objective is to show that i) the trip-specific sequences are significantly tighter than the population-intervals, at the same coverage level, for short and long trips, and ii) adding higher-order correlation terms to~\eqref{eq:sample-variance-residual} do not necessarily lead to significantly tighter sequences, at least not in this study.

\begin{figure}[!t]
  \centering
\captionsetup[subfigure]{labelformat=empty}
  \subfloat[][]{\includegraphics[width=1\textwidth]{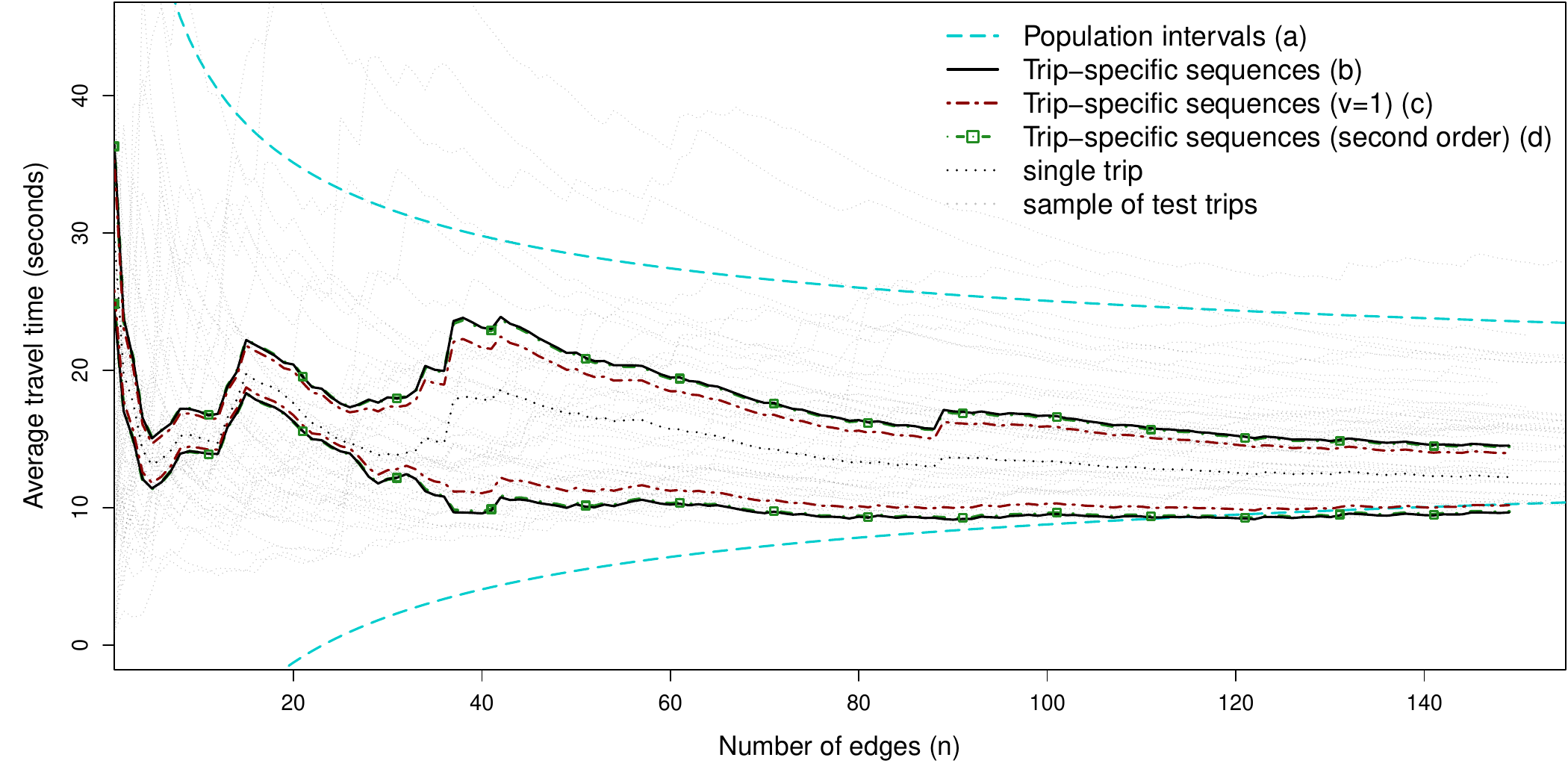}}
  
    \subfloat[][]{\includegraphics[width=0.6\textwidth]{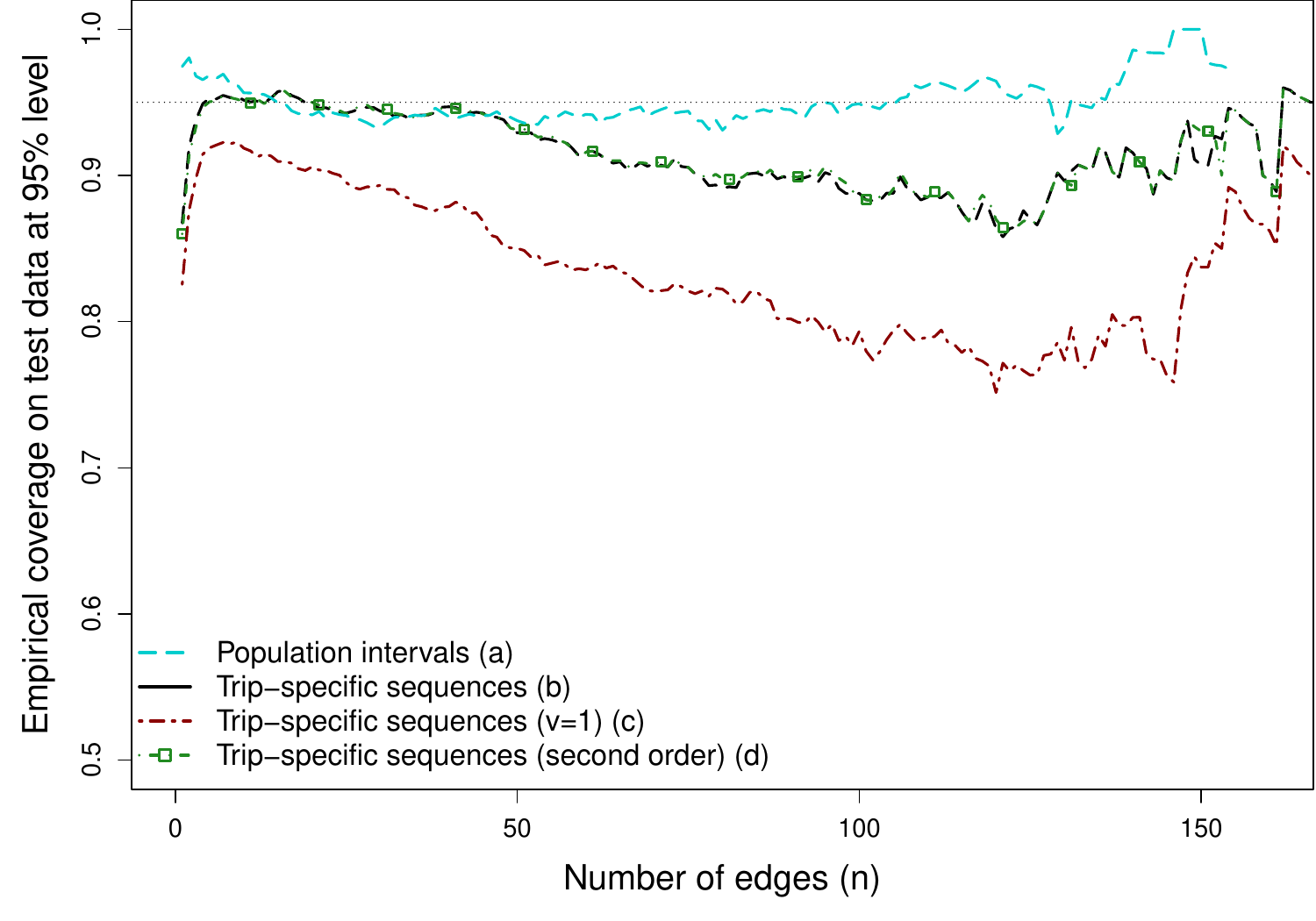}} 
   \subfloat[][]{\includegraphics[width=0.42\textwidth]{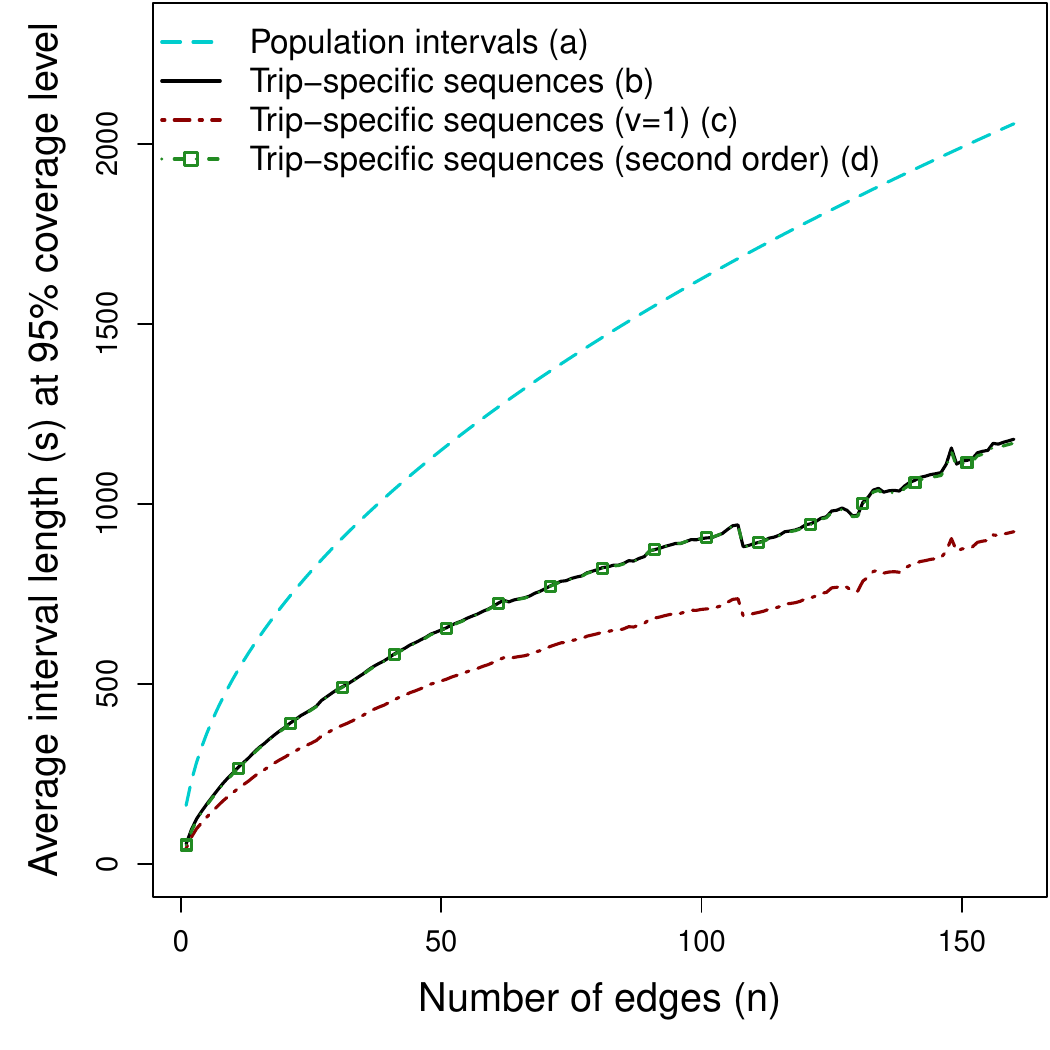}}
  \vspace{-1\baselineskip}
  \caption{{ (Top) Compassion between the (a) population prediction intervals in~\eqref{eq:prediction-interval}, to (b) trip-specific prediction sequences~\eqref{eq:prediction-sequence} for an arbitrary trip of 149 edges (dotted black), to (c) trip-specific sequences with \(\hatnu=1\) in~\eqref{eq:sample-variance-residual}, and to (d) trip-specific sequences calculated by adding a second-order correlation to the variance sequence in~\eqref{eq:estimator-sigma-reduced} as in~\eqref{eq:second-order-term}. The progressive average \((n^{-1}\T_{\path})\) of 500 arbitrary test trips is in dotted grey. Empirical coverage and interval length of 2,000 test trips are illustrated in bottom left and right respectively. Coverage is calculated at the theoretical 95\% significance level. 
  }}
  \label{fig:comparision-of-prediction-intervals}     
\end{figure}

Figure~\ref{fig:comparision-of-prediction-intervals} ({ top}) illustrates (a) the population prediction intervals~\eqref{eq:prediction-interval} for the sampled-at-random approach, (b) the trip-specific prediction sequences~\eqref{eq:prediction-sequence} for an arbitrary trip of 149 edges starting at 6:40 a.m., and travelling for 35.1~km over a period of 30 minutes, (c) the trip-specific sequences while setting the sample variance of residual to one (i.e.~\(\hatnu=1\)  in~\eqref{eq:sample-variance-residual}), and (d) the trip-specific sequences calculated by adding a second-order correlation to the variance sequence in~\eqref{eq:estimator-sigma-reduced}, as
\begin{equation}
\label{eq:second-order-term}
\hatsigma_{\path}(\Tstar)  +  2\hatxi^{(2)} \sum_{\langle e, \tilde e, e'\rangle \in \path}d_{e}d_{e'}\hatsd_{e}(\Tstar) \hatsd_{e'}(\Tstar),
\end{equation}
where \(\hatxi^{(2)}\) is calculated as~\eqref{eq:xi-estimate}, although for the correlation between endpoints of every 3 edge sequences \(\langle e, \tilde e, e'\rangle\), instead of \(\langle e, e'\rangle\). For each of (a), (b), (c) and (d), Figure~\ref{fig:comparision-of-prediction-intervals} { (bottom left)} illustrates the empirical coverage levels at the theoretical 95\% levels for each length \(n\), for the 2,000 trips of the test data, and the progressive averages (\(n^{-1}\T_{\path}\)) (in grey) for 500 arbitrary selected trips. Fewer than 43 out of 2,000 test trips travelled more than 150 edges, therefore we limited the x-axis of the figure to 155. { The average interval width of the prediction intervals (a), (b), (c) and (d) are illustrated in Figure~\ref{fig:comparision-of-prediction-intervals} (bottom right).}

All intervals and sequences are calculated using the corresponding data in Table~\ref{tb:parameter-estimates}. All test trips share the same population intervals. Prediction sequences~\eqref{eq:prediction-sequence} are constructed for each test trip. Empirical coverage is calculated as the average number of test trips with travel time between the given intervals/sequences, at each length \(n\). 

Trip-specific sequences (b), (c) and (d) result in tighter intervals than the prediction intervals (a), approximately half as tight. { The tightness of the trip-specific sequences is especially evident for the first 40 edges of the trip.} The empirical coverage level matches the theoretical 95\% level of significance for almost the whole range for all intervals except (c) when setting \(\hatnu =1\), which leads to slightly tighter sequences than (b), where \(\hatnu = 1.30\), but does not attain the required coverage. 

The integration of \(\hatnu\), as a correction scalar to the variance in~\eqref{eq:prediction-sequence}, improves the empirical coverage probability of the trip-specific prediction sequences (b) across the whole range, attaining the theoretical 95\% coverage level. In (d), the second-order correlation coefficient is estimated to be \(\hatxi^{(2)} = 0.16\), which leads to a residual variance estimate of \(\hatnu = 1.23\), smaller than that of (b) (1.3). This is expected since \(\hatxi^{(2)}\) accounts for extra variability in the data. { However, adding the second-order correlation in~\eqref{eq:second-order-term} did not lead to a significant reduction in the width of the prediction sequences, nor a significant improvement in coverage probabilities, in comparison to (b). As shown in Figure~\ref{fig:comparision-of-prediction-intervals}, lines (b) and (d) almost fully overlap}. Very few trips travel over many edges (circled-solid line in the top right panel of Fig.~\ref{fig:coverage-alternative-models}), and this contributes to the variability of empirical coverage at a higher number of edges.

Table \ref{tb:numerical-results-sequence} illustrates various numerical results for the test data used in Figure~\ref{fig:comparision-of-prediction-intervals}, for population prediction intervals (a) and sequences (b). The integration of \(\hatnu\) improves the empirical coverage probability of the trip-specific prediction sequences by approximately 8 percentage points, for the sampled-at-random approach, and 10 points for the AM-rush stratum. 
\begin{table}[!htp]
  \caption{Model assessment for the trip-specific predicting sequences (PS) and the population prediction intervals (PI) methods under different sampling methods. Numerical results associated with PS are listed for \(N(0,\hatnu)\) and \(N(0,1)\), separated by a comma, all metrics are in seconds, if not a percentage}
  \label{tb:numerical-results-sequence}
  \centering
  {\footnotesize
    \begin{tabular}{l  SSSSSSS}
                                     & \multicolumn{2}{c}{At random} &        & \multicolumn{4}{c}{Stratified sampling by traffic bins}                        \\
                                     &                               &        & \multicolumn{2}{c}{Trip-specific PS} & \multicolumn{2}{c}{Population PI}           \\
                                     & {Trip-specific PS}                              & {Population PI}        & {AM-rush}                            & {Non-rush}     & {AM-rush} & {Non-rush} \\[5pt]
      Root mean squared error        & 242.2                         & 379.9  & 237.8                                & 260.8          & 383.4     & 288.1      \\
      Mean absolute error            & 167.6                         & 285.1  & 168.0                                & 183.0          & 289.8     & 194.3      \\
      Mean error                     & -1.9                          & -17.7  & -3.9                                 & 6.5            & -47.4     & -6.2       \\
      Mean absolute \% error & 14.4                          & 26.8   & 14.4                                 & 15.0           & 26.6      & 23.4       \\    
      Empirical coverage~(\%)        & {91.7, 84.2}                  & 94.2   & {94.4, 84.1}                         & {86.5, 83.3}   & 97.8      & 95.6       \\
      Interval length                & {747, 584}                & 1388 & {829, 569}                      & {646, 590} & 1760    & 1022     \\
      Interval relative length~(\%)  & {71.4, 55.8}                  & 140.5  & {79.6, 54.6}                         & {60.4, 55.2}   & 167.9     & 137.1 
  \end{tabular}
  }
\end{table}

The relative length of the prediction sequences (b) to the trip's travel time has dropped significantly in comparison to the prediction intervals. For the sampled-at-random approach, the relative length is 71.4\%, almost half of (a) at 140.5\%. This reduction ratio is consistent for different sampling methods. Other metrics also improved. For example, the predictive mean error dropped from -17.7 for (a) to -1.9 seconds, for the sampled-at-random. This drop is consistent across all sampling strata, except the Non-rush stratum. Most edges travelled in the Non-rush stratum have very few observations, unlike rush hour strata. Hence, they have been imputed by time-bin estimates, see Section~\ref{sec:traff-bin-estim} for more details on this imputation. {It is expected that the trip-specific mean ($\mu_\path(t_0))$ leads to less predictive bias than its counterpart population version in~\eqref{eq:sample-mean}, since the former includes route-level information.}

As established in Theorem~\ref{thm:clt-traveltime-offline}, the prediction intervals~(\ref{eq:prediction-interval}), when constructed for the average (\(n^{-1}\Tnew_{\path}\)), converges to zero theoretically as \(n\) increases. This is not the case for the trip-specific sequences~\eqref{eq:prediction-sequence}. We illustrate the empirical shrinkage of the predictive intervals in SM Table~\ref{app:tb:numerical-results-n}, which reports model performance under different trip lengths of test data. In summary, while the empirical coverage probability sustains the theoretical level of 95\%, the average interval length drops to 92.2\% of the observed travel time for trips with \(n>120\), in comparison to 242\% for trips with \(n<40\). Such asymptotic shrinkage is feasible for applications with very long trips. For example, using the sampled-at-random estimates in Table~\ref{tb:numerical-results-sequence}, only the top three trips in the number of edges (out of 19,967) have trip-specific sequences wider than the population intervals. Those three trips travel at least 305 edges over a distance of at least 77~km. SM Figure~\ref{app:fig:trip-hist-prediction-sequence} illustrates the distributional fit of the trip-specific sequences of the test data to a standard normal. That is to say, it plots the left-hand side of~\eqref{eq:predictive-clt} to a N(0,1).

\subsection{Comparison to alternative models}\label{sec:comp-altern-models}
On the same out-of-sample test data, we compare our proposed trip-specific sequences to alternative models that focus on travel uncertainty, with emphasis on empirical coverage levels, length of coverage intervals, and estimation bias. 

\cite{woodard2017predicting} proposed a generative model based on log-normal mixtures for the distribution of speed, with edge-specific states representing congestion. They use a Hidden Markov chain model (HMM) to estimate congestion states and account for other sources of dependency by augmenting the log-normal mixture with a trip-specific random effect. We refer to this model as {\it HMM+Trip-effect}, and implement it with 2 hidden congestion states. We also implement a variant of the HMM+Trip-effect that assumes no within-trip dependency and no random effect, as a simple sum of independent log-normals. We refer to this model as the no-dependence model. Prediction intervals for HMM+Trip-effect and no-dependence models are calculated as in~\cite[Algo. 2]{woodard2017predicting}. In particular, for each new trip, we sample 1,000 travel times for the first edge at the start-time traffic-bin, and iteratively, for each of the 1,000 samples, we sample a travel time of the second edge at the traffic bin of the start-time plus the travel time of the first edge, and so on until the last edge. The predictive intervals are then the empirical intervals of those 1,000 samples of total travel time, and the prediction is the arithmetic mean of those samples.

We also compare our proposed intervals to a regression-based approach that models travel time \citep{budge2010empirical, westgate2013travel}. In our case, we use a standard linear regression model with the log of travel time as a response variable and total route distance and the traffic-bin of the trip's start time (categorical) as predictors. The assumptions of the linear regression model hold approximately in QCD.

The 17,967 trips of the training data are used to estimate the parameters of all models. Trip-specific prediction sequences are calculated as in Section~\ref{sec:pool-based-prediction-intervals}, with parameters in Table~\ref{tb:parameter-estimates}. Parameters of the HMM+Trip-effect and no-dependence models are calculated following \citet[Algo. 2]{woodard2017predicting}, which we implemented in an R-package\footnote{\texttt{melmasri.github.io/traveltimeHMM}}.

\begin{figure}[!htp]
  \centering
  \captionsetup[subfigure]{labelformat=empty}
 \subfloat[][]{\includegraphics[width=0.49\textwidth]{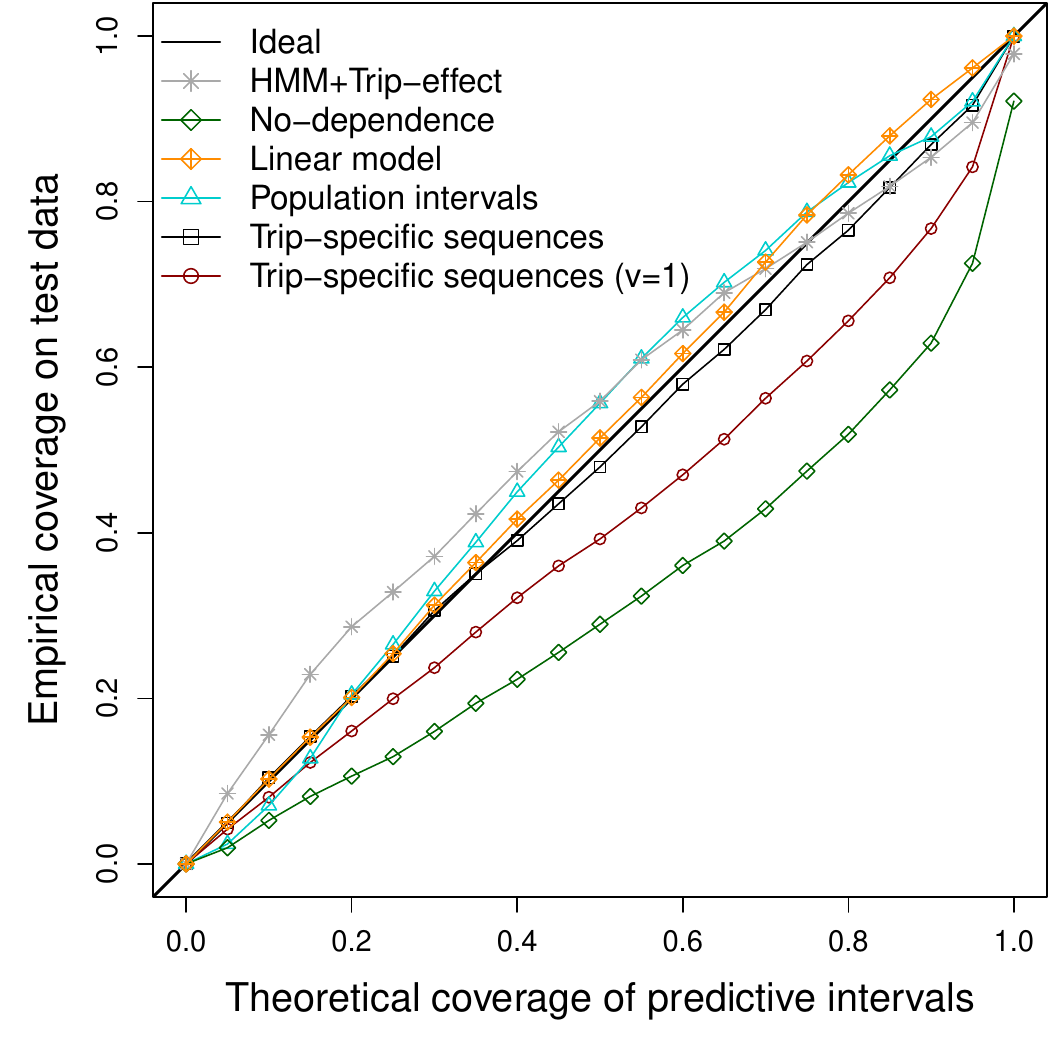}}
    \subfloat[][]{\includegraphics[width=0.49\textwidth]{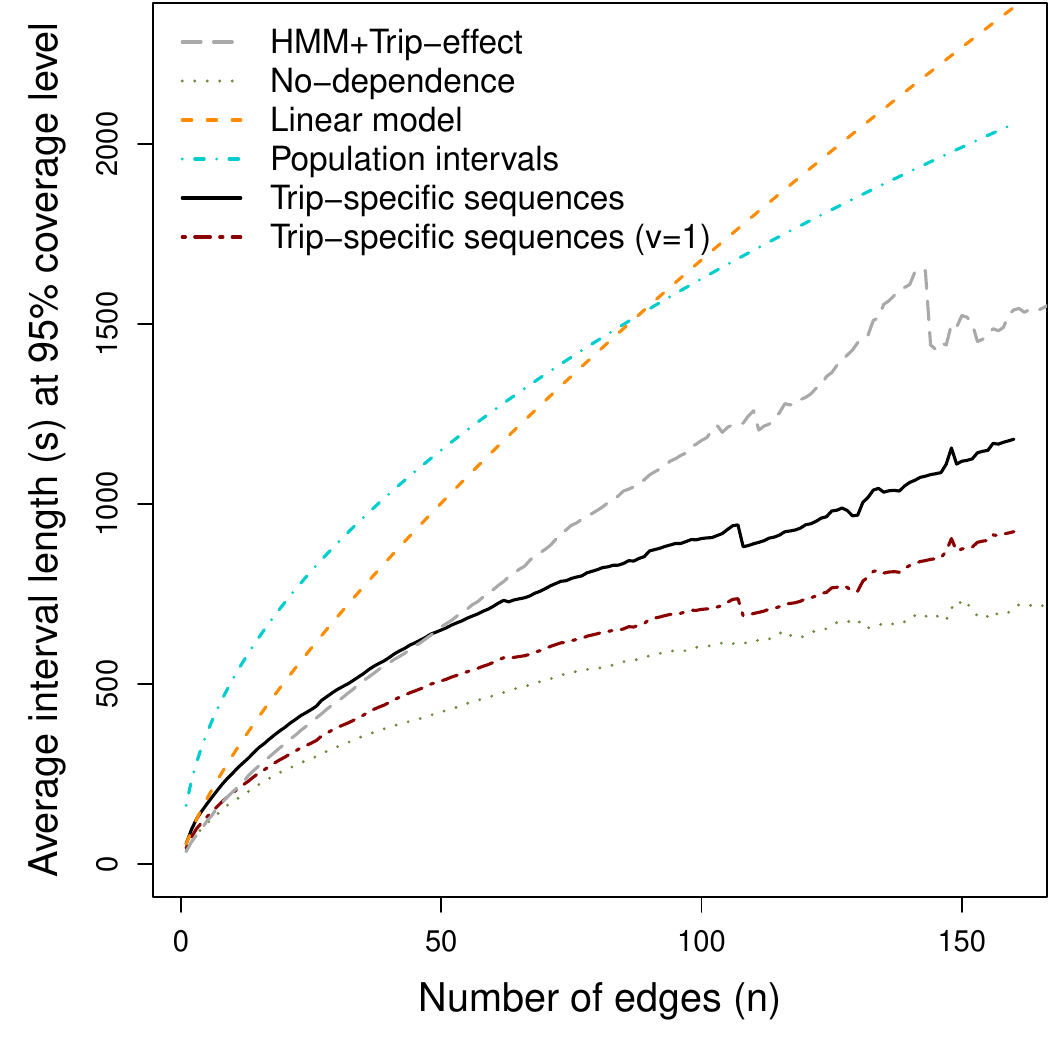}} 
 
  \subfloat[][]{\includegraphics[width=0.6\textwidth]{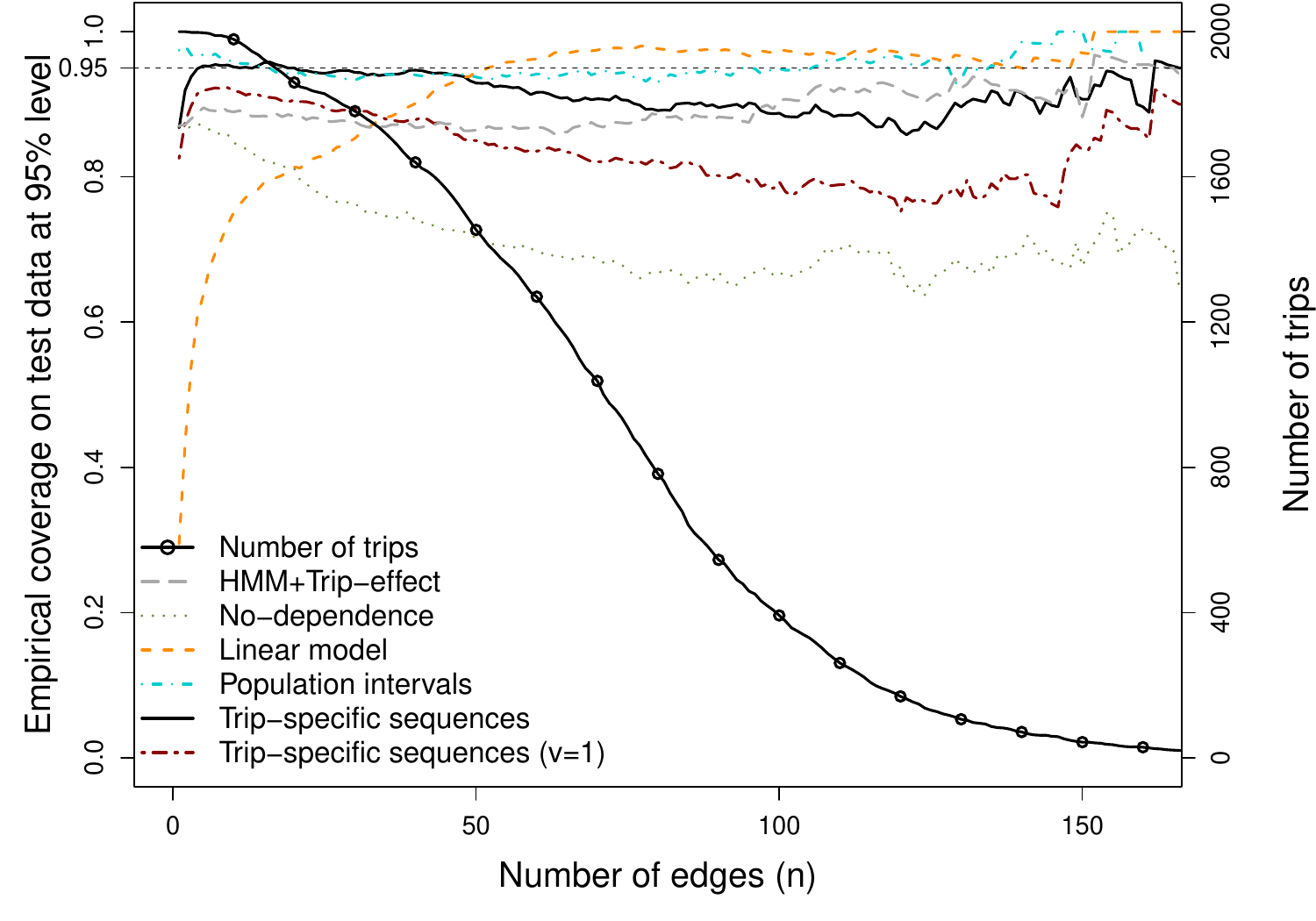}}

  \vspace{-1\baselineskip}
  \caption{Comparison of empirical coverage against theoretical levels (left) and distance (bottom). The right figure illustrates the width of prediction intervals against distance.}
  \label{fig:coverage-alternative-models}
\end{figure}

Figure~\ref{fig:coverage-alternative-models} illustrates empirical coverage results, for the 2,000 test trips, against theoretical levels (left panel) and the number of travelled edges \((n)\) (bottom panel). Our proposed trip-specific sequences~(\ref{eq:prediction-sequence}) and population intervals~(\ref{eq:prediction-interval}) both sustain the theoretical levels, as does the linear model. The HMM+Trip-effect model achieves the theoretical level only at higher levels. Our proposed intervals also sustain the theoretical 95\% coverage levels across distance, with slight variability resulting from the strong drop in the number of trips having many edges. The average length of an edge is 170 meters, making the average 50-edge trip around 8.5~km. The average interval width of our trip-specific sequences is 855 seconds at empirical coverage of 94.8\%, while the HMM+Trip-effect is at 870 seconds with 90\% coverage. The average interval width of our trip-specific sequence grows sub-linearly with distance, with a slope parallel to the no-dependence model and slower than alternative models, as shown in the right panel of Figure~\ref{fig:coverage-alternative-models}. 

By accounting for the rotation map~\eqref{eq:recursive-map}, our trip-specific sequences achieve negligible predictive bias (-0.7 seconds) in comparison to alternative models, as shown in Table~\ref{tb:model-comparison}, which illustrates numerical results for the 2,000 test trips. This is not surprising, given that the recursive mean estimate~(\ref{eq:mu-recursive}) hinders the accumulation of bias with distance. This also resulted in a significant reduction of mean absolute and squared errors of our trip-specific sequences in comparison to the alternatives. The generative predictive sampling process of \cite[Algo. 2]{woodard2017predicting} also resembles the rotation map in~(\ref{eq:recursive-map}); however, unlike \cite{woodard2017predicting}, we do not assume a distribution for speed or a specific form of serial dependency.
\setlength\tabcolsep{4pt}
\renewcommand{\arraystretch}{0.85}
  \begin{table}
    \caption{Comparing trip-specific prediction sequences to alternative models}
    \label{tb:model-comparison}
    {\footnotesize
    \begin{tabular}{l SSSSS} 
                                              & {Trip-specific sequences} & {HMM+Trip-effect} & {No-dependence} & {Linear model} \\[5pt]
          Root mean squared error             & 239.0              & 309.9             & 283.4           & 364.9          \\
          Mean absolute error                 & 167.6              & 207.1             & 187.1           & 266.2          \\
          Mean error                          & -0.7               & -39.5             & 40.1           & 18.2           \\
          Mean absolute percentage error~(\%) & 14.5               & 17.7             & 15.5           & 24.3           \\
    \end{tabular}
    }
  \end{table}

  Even though the HMM+Trip-effect improved the coverage probability in comparison to the no-dependence model, they also reduce prediction accuracy. The mean absolute percentage error for HMM-Trip-effect is 17.7\%, while for the simpler no-dependence model it is 15.5\%. This pattern is consistent with the results of \citet[Table 1]{woodard2017predicting}. This, alongside the low transition probability in \citet[Fig. 6]{woodard2017predicting}, suggests that travel time is serially dependent but not necessarily Markovian, motivating our general mixing approach.

  Our trip-specific sequences (i) have fewer parameters than the HMM+Trip-effect, (ii) do not require a generative sampling method for estimation, and more importantly (iii) do not require vehicle history to calculate a trip-specific random effect, while (i) and (ii) are required for HMM+Trip-effect. Trip-specific sequences use \(|E|\times\text{\# time bins}  + 2\) (for the \(\hatnu\) and \(\hatxi\)) parameters, while HMM+Trip-effect uses \(|E|\times \text{\# time bins} \times \text{\# hidden states} + 1\), at least double the trip-specific sequences; the extra 1 is for the variance of the random effect. Because of (i), (ii) and (iii), our approach is computationally efficient and hence reliable for large-scale implementations.

  The no-dependence model is conceptually similar to our trip-specific sequences, in the number of parameters and approach as the sum of independent log-normal random variable, in a sense assuming \(\hatxi=0\) in~\eqref{eq:estimator-sigma-reduced}.

  \section{Discussion}\label{sec:discussion}
   Our results build on the assumption that the distribution of speed over road segments has a periodic mean and covariance function (cyclostationary) with respect to time. Under such an assumption, we establish the normality of the ratio of travel time to distance. This suggests that the empirically observed~\citep{woodard2017predicting,guo2012multistate} log-normality of travel time is an artifact of the network topology, i.e.~the distribution of distance influenced by urban planning. By conditioning on distance, travel time is at most a mixture of normals. {By conditioning on start/end locations, travel time is a mixture of a mixture of normals, here the first mixture is over possible routes.} 
  
 With such observations, it is not surprising that regression-based models have shown promising results in travel time modelling \citep{westgate2016large, woodard2017predicting,budge2010empirical, westgate2013travel}. In particular, our work suggests a Gaussian form as a population-style distribution for travel time, where both the mean and variance scale with distance. For example, an \(N(n\mu, n\sigma^{2})\), where \(n\) is the number of road segments and \((\mu, \sigma^{2})\) are map- (possibly traffic bin-) specific mean and variance constants. A trip-specific model is of the form \(N(\mu_{i}, \sigma_{i}^{2})\), where \(\mu_{i}\) is the \(i\)th trip mean, as the sum of average travel times of the \(n\) segments of the route with \(\sigma_{i}^{2}\) being the covariance of the variables of the sum.

We develop reliable and computationally efficient inference methods to estimate the parameters of such population and trip-specific distributions of travel time. Our methods rely on the first- and second-moment estimates of speed distribution on edges of the network, which do not require any complex estimation methods, resulting in analytical prediction intervals that are interpretable, require minimal computational complexity and attain the theoretical coverage levels. Our trip-specific prediction intervals are suitable for short and long trips, providing tighter bounds than competing models (Fig.~\ref{fig:comparision-of-prediction-intervals}). Since they are composed of the sum of second-moment estimates of speed on each road, they are suitable for high-throughput, low-latency applications. We implemented our method in an \texttt{R}-package, located at~\texttt{https://github.com/melmasri/traveltimeCLT}, and implemented some competing models at~\texttt{https://melmasri.github.io/traveltimeHMM}.
  
  The effectiveness of our trip-specific intervals is a result, first, of our insight into the long-term Gaussian-based distribution of travel time, and second, of accounting for the dependency of speed on time, which helps in reducing the predictive error that accumulates with distance to an almost negligible bias. This bias accumulates by summing the estimation biases on each segment on a route. 
  
  Without cyclostationarity, our central limit theorem results (Thm.~\ref{thm:predictive-traveltime}, Lem.~\ref{thm:slln-uniform}) still hold, though only for parameters that depend on initial conditions and route. In other words, we require real-time data to be able to adjust the model parameters in an online manner as the ride progresses. Inference for such parameters can be carried \mo{out}, for example, by a blocking method \citep{wu2009recursive, peligard1995estimation} if a large enough segment of a progressing trip is provided. However, theoretical properties of such estimators are difficult to derive and generalize due to contamination from unpredictability of real-time events. Online routing systems\footnote{the likes of Google Maps.} generate their travel estimates in two stages, a coarse ETA estimate topped up by a real-time correction. The coarse estimate represents travel time under stable traffic conditions, captured in the traffic training data. The real-time model corrects for sudden changes in traffic or slight deviations from the captured states of stationarity. Here, we focused on understanding the properties of travel time under stable traffic conditions. In that sense, the traffic training data used to estimate the average and variance of speed on edges of the network should to some extent reflect traffic states a future trip will drive through. Frequent updates of traffic data are required to integrate local changes in traffic, e.g.~to integrate closures, detours, and prolonged weather conditions.
  
  
  Our approach enables further statistical and applied research on such topics, with many open questions, including: Given a distribution of distance, how can the limit distributions be used to simultaneously sample routes and travel time to retrieve network dynamics mimicking that of the initial input? How can multiple route variances be pooled to construct an efficient test statistic for the difference of percolation regimes, i.e.~travel times? How can the hypothesis that travel time on a route is faster and/or less variable than on another route be tested efficiently? 
   
  \section*{Acknowledgements} We would like to thank Joshua Stipancic for providing a cleaned version of the data, \'Eric Germain and Adrien Hernandez for providing support for code development. ME gratefully acknowledges the Natural Sciences and Engineering Research Council of Canada (NSERC) PDF and the Institute for Data Valorisation (IVADO) for the funding they provided. A majority of this work was conducted at the Department of Decision Sciences, HEC Montr\'eal and the Mila Quebec Artificial Intelligence Institute. We thank the referees for their comments and suggestions that helped us improve the article.
 

\bibliographystyle{imsart-nameyear}
\bibliography{ref.bib}

\begin{appendices}
\setcounter{section}{0}
\setcounter{figure}{0}
\setcounter{table}{0}
\setcounter{equation}{0}
\renewcommand{\thesection}{A\arabic{section}}  
\renewcommand{\thefigure}{A\arabic{figure}}
\section{Detailed algorithm for trip-specific intervals}
Here, we present a summary of how to construct and estimate the required parameters for the trip-specific sequences in Section~\ref{sec:predictive-distribution}.

\noindent\begin{minipage}{\textwidth}
\renewcommand\footnoterule{}    
\begin{algorithm}[H]
\caption{Trip-specific intervals}\label{algo:trip-specific}
\begin{algorithmic}
\Require Given a set of traffic data, compute \((\hatm_{e}, \hatsigma_{e}, e \in E)\) as the sample mean and variance, respectively, for speed observations of every edge and time bin. For example, a total of $168 \times |E|$ estimates, if hour of week time bins are used\footnote{All calculations described are based on the time bin into which the starting time falls. For example, an observed speed at start time $t$ is assigned $\hatmu_e(t)$ of the time bin into which $t$ falls.}.
\Require A set of $M$ training trips, of observed travel times $\T_\path ^{(j)}, j=1, \dots, M$.
\State Compute $\hatxi$ as in~\eqref{eq:xi-estimate}.
\For{$j = 1, \dots, M$}
    \State For $k=0, \dots, K$ edges of the $j$th trip, and start time $t_0$, compute \(t_{k+1} = t_{k} + d_{k} \hatm_{k}({t_{k}})\).
    \State Compute $\hatmu ^{(j)}_{\path}(t_0)$ of~\eqref{eq:estimator-mu-recursive} as \(\hatmu ^{(j)}_{\path}(t_0) = \sum_{k=0}^K d_k \hatm_k(t_k)\).
    \State Compute \(\hatsd^{(j)}_{\path}(t_0)\) of~\eqref{eq:estimator-sigma-reduced} as
        \[(\hatsd^{(j)}_{\path}(t_0))^2= \sum_{k=0}^K d_{k}^{2}\hatsigma_{k}(t_k) + 2\hatxi \sum_{k=1}^K d_{k}d_{k-1}\hatsd_{k}(t_k)\hatsd_{k-1}(t_{k-1}).\]
    \State Compute $\varepsilon^{(j)}$ as \(\varepsilon^{(j)} = \big(\hatsd^{(j)}_{\path}(t_0) \big)^{-1}  \big( \T_{\path}^{(j)} - \hatmu_{\path}^{(j)}(t_0) \big) .\)
\EndFor

\State Compute $\bar \varepsilon$ as \(\bar \varepsilon = {M}^{-1}\sum_{j=1}^{M}\varepsilon^{(j)}.\)

\State Compute $\hatnu^2$ as in~\eqref{eq:sample-variance-residual}, that is
    \(\hatnu^{2} = {(M-1)}^{-1} \sum_{j=1}^{M}(\varepsilon^{(j)} - \bar \varepsilon)^{2}.\)
    
\State Given a new trip $\Tnew_{\path'}$, with start time $t'_0$, over route $\path'$ over $K'$ edges, compute \( \hatmu_{\path'}(t'_0)\), and  \((\hatsd_{\path'}(t'_0))^2\) following a step from the for loop above. Prediction intervals for  $\Tnew_{\path'}$ follow from~\eqref{eq:prediction-sequence}, as
\[\Tnew_{\path'} \in \Bigg [\hatmu_{\path'}(t'_0) - z_{\qparam/2}\sqrt{\hatnu^{2} \hatsd^{2}_{\path'}(t'_0)},\quad \hatmu_{\path'}(t'_0) + z_{1-\qparam/2}\sqrt{\hatnu^{2} \hatsd^{2}_{\path'}(t'_0)} \,\Bigg ].\]
\end{algorithmic}
\end{algorithm}
\vspace{-1cm}
\end{minipage}

\section{Empirical ergodicity and parameter estimation} \label{app:empirical-ergodicity}
\begin{figure}[!htbp]
    \centering
  \subfloat[][]{\includegraphics[width=0.5\textwidth]{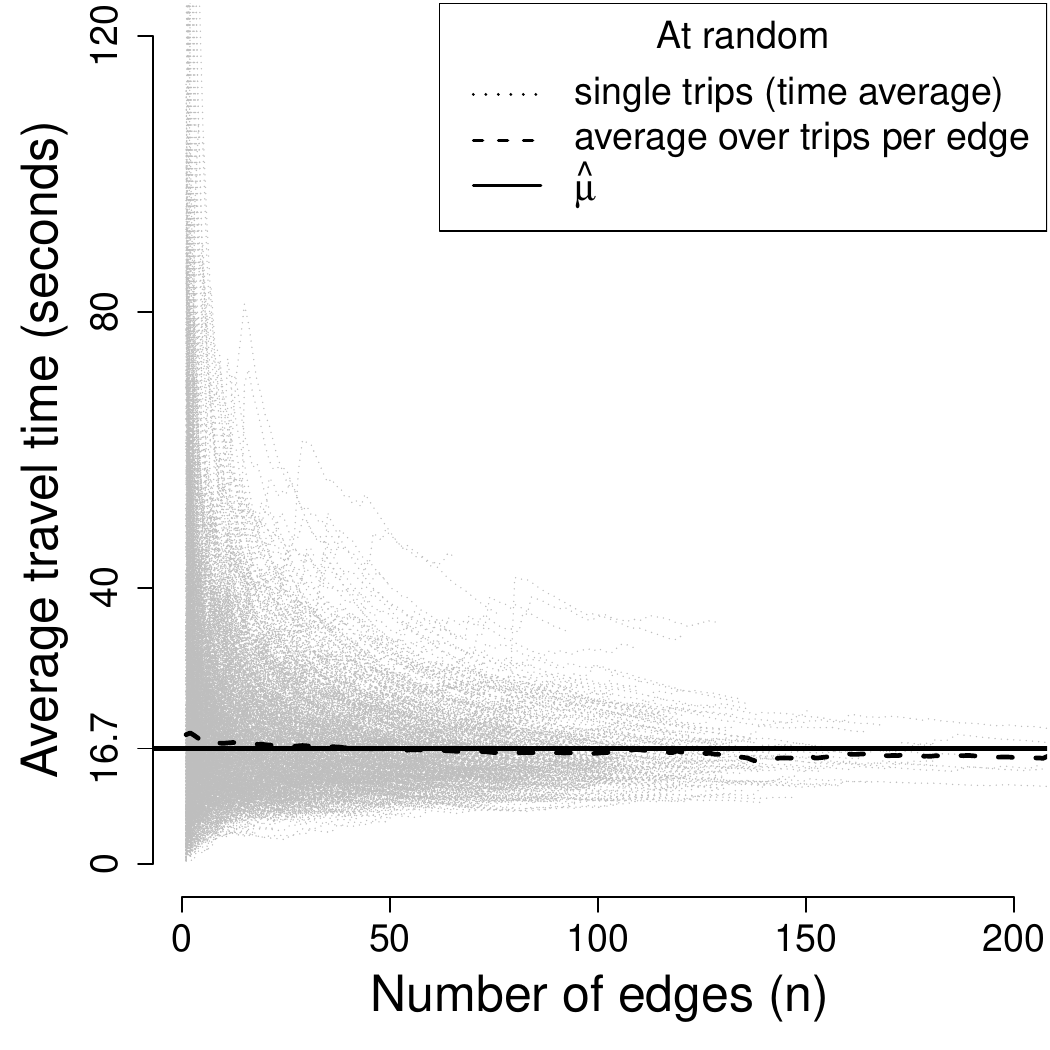}}
  \subfloat[][]{\includegraphics[width=0.5\textwidth]{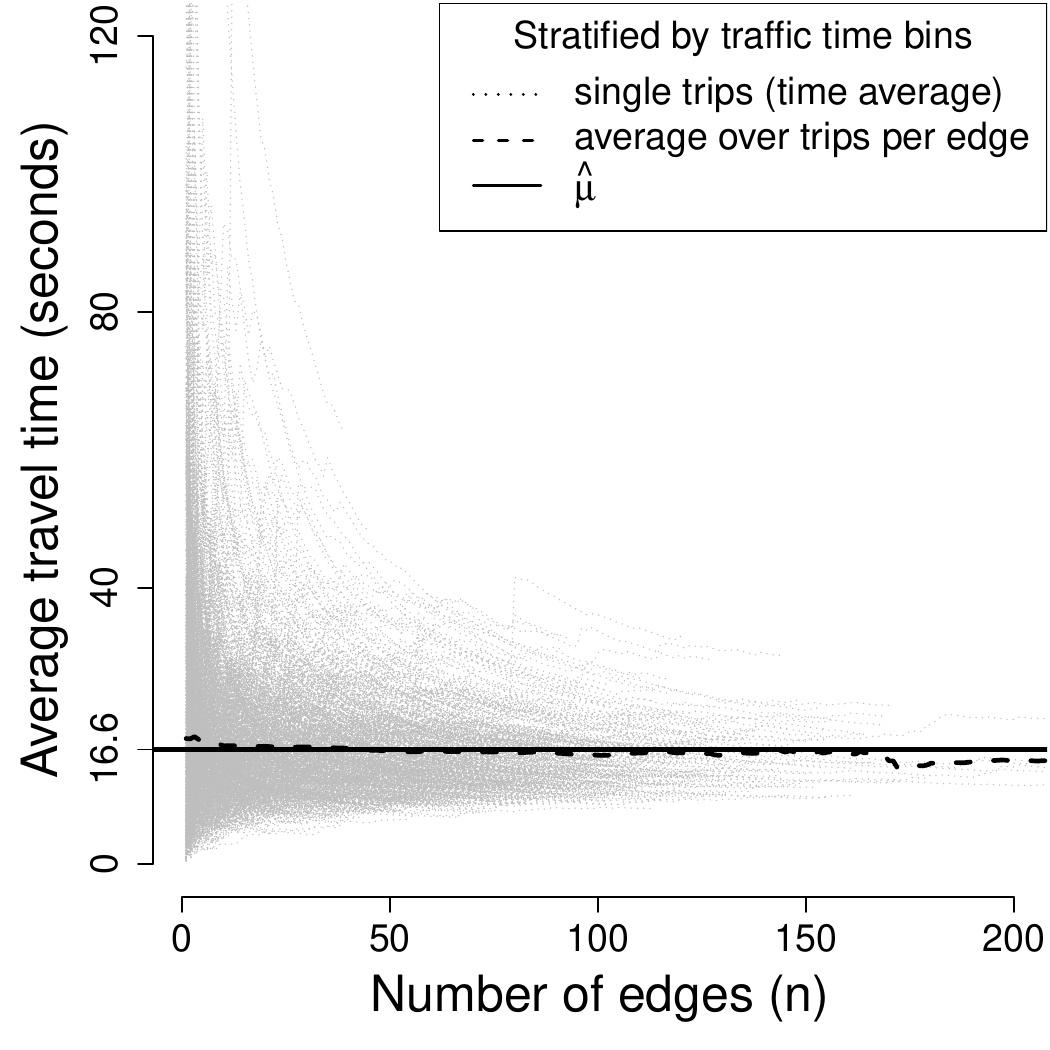}}
  \vspace{-1\baselineskip}
  \caption{Time and space averaging for 1,000 trips that are sampled-at-random (left), and for the 1,500 trips stratified by traffic-bins (overall estimation, right). All trips are of at least 10 edges. The time average of each trip is in grey dotted lines, dashed lines represent (space) averaging over trips per length and solid lines are the estimates \(\hatmu\).}
  \label{fig:hatmu-different-sampling-methods}
\end{figure}

To illustrate the empirical ergodicity of the system, Figure \ref{fig:hatmu-different-sampling-methods} reports the space average (solid and dashed lines) as the estimate of $\E[k^{-1}T_\path \mid k = n]$ by averaging the average travel time for the first $n$ edges of each trip, across multiple trips, for each length $n$, and for whole trips as in $\hatmu$. The figure also reports the time average as $n^{-1}\T_\path$ for each trip (dotted lines), for the sampled-at-random trips (left) and by traffic-bins (right). The space average is exactly the same, whether calculated as an average per length $n$, or over all trips. The time average of very long trips converge to the space average, indicating the ergodicity of the system.
\end{appendices}

\clearpage
\begin{supplement}
  \input{Supplementary}
\end{supplement}


\end{document}

%% file: myprem.tex
\usepackage{amsmath}
\usepackage{amssymb}
\usepackage{times}
\usepackage{bm}
\usepackage{multirow}
\usepackage{booktabs}
\usepackage{graphicx}
\usepackage[colorlinks=true,citecolor=blue,urlcolor=blue]{hyperref}
\usepackage{subfig}
\usepackage{siunitx}
\usepackage[english]{babel}
\usepackage{xr}
\usepackage{algorithm}
\usepackage{algpseudocode}
\usepackage{bbm}

\usepackage{amsthm}
\usepackage[utf8]{inputenc}
\usepackage[T1]{fontenc}
\usepackage{graphicx}

\usepackage[round, sort&compress]{natbib}   
\usepackage[toc,page]{appendix}

\graphicspath{{./img/}}
\def\trho{{\rho^*}}
\def\R{\mathbb{R}}
\def\F{\mathcal{F}}
\def\borelset{\mathcal{B}}
\def\path{\rho}

\def\Q{\mathbb{Q}}
\def\T{\mathcal{T}}
\def\Z{\mathbb{Z}}
\def\Tnew{\T^{\text{new}}}
\def\bT{{\boldsymbol T}}
\def\Var{\mathbb{V}}
\def\Corr{{\text{Corr}}}

\def\L{\mathcal {L}}
\def\qparam{\beta}
\def\E{{\mathbb{E}}}
\def\P{{\mathbb{P}}}
\def\hatmu{{\hat \mu}}
\def\hatm{{\hat m}}
\def\hatsigma{{\hat \sigma^2}}
\def\hatsd{{\hat \sigma}}
\def\xiprof{\xi_\text{prof}}
\def\hatxi{{\hat{\xi}_\text{prof}}}

\def\hatnu{{\hat \nu}}
\def\sigmaprofile{\hat \sigma_\text{prof}}

\def\sighatmu{{\widehat \Var(n^{-1}\T_{\path})}}
\def\hatn{{\widehat \E[n^{-1}]}}
\def\spi{\pi}

\def\Tstar{t^*}
\newcommand\independent{\protect\mathpalette{\protect\independenT}{\perp}}
\def\independenT#1#2{\mathrel{\rlap{$#1#2$}\mkern2mu{#1#2}}}

\newtheorem{theorem}{Theorem}
\newtheorem{lemma}[theorem]{Lemma}

\newtheorem{definition}[theorem]{Definition}
\newtheorem{example}[theorem]{Example}

%% file: Supplementary.tex


\setcounter{section}{0}
\setcounter{figure}{0}
\setcounter{table}{0}
\setcounter{equation}{0}
\renewcommand{\thesection}{S\arabic{section}}  
\renewcommand{\thetable}{S\arabic{table}}  
\renewcommand{\thefigure}{S\arabic{figure}}
\renewcommand{\theequation}{S\arabic{equation}.\arabic{equation}} 
\section{Additional data results} \label{app:within-across-trips}

Table~\ref{app:tb:numerical-results-n} illustrates coverage interval metrics in relation to trip length \((n)\).
\begin{table}[ht!]
  \caption{Model assessment for the asymptotic method for trips with different lengths (sampled at random).}
  \label{app:tb:numerical-results-n}
  \centering
  {\footnotesize
  \begin{tabular}{l SSSS} 
    & \(n \leq 40\) & \(40 < n \leq 80\) & \(80 <n  \leq 120\) & \(n > 120\) \\
   Mean absolute percentage error~(\%)                  &  34.84  &   26.40 &  24.77  &   21.48 \\ 
   Empirical coverage~(\%)                  &  95.00&    95.00 &  94.60&    96.00\\ 
   PI length as percentage trips duration~(\%)        & 242.63 &  137.51 & 110.23&    92.21\\ 
  \end{tabular}
  }
\end{table}

\begin{figure}[!htbp]
  \centering
  \includegraphics[width=0.5\textwidth]{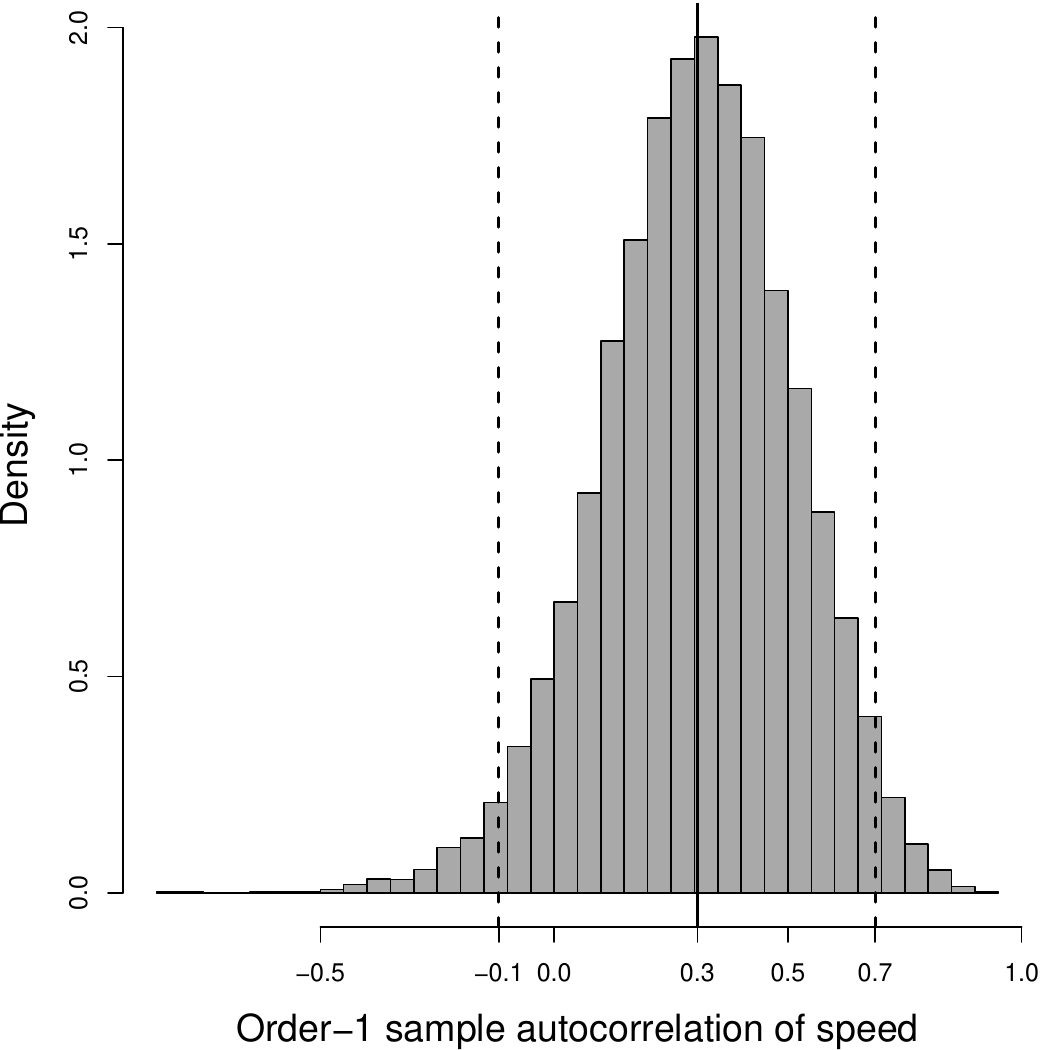}
  \caption{Histogram of \(\xi_{G}\) in (\ref{eq:xi-estimate}), of 17,967 trips in the training data described in Section \ref{sec:comp-altern-models}. The mean is indicated with black dashed lines, and the 95\% empirical confidence intervals are in dashed red.}
  \label{app:fig:lag-1-autocorrelation}
\end{figure}

  
  

\begin{figure}[!htbp]
  \centering
  \subfloat[][]{\includegraphics[width=0.5\textwidth]{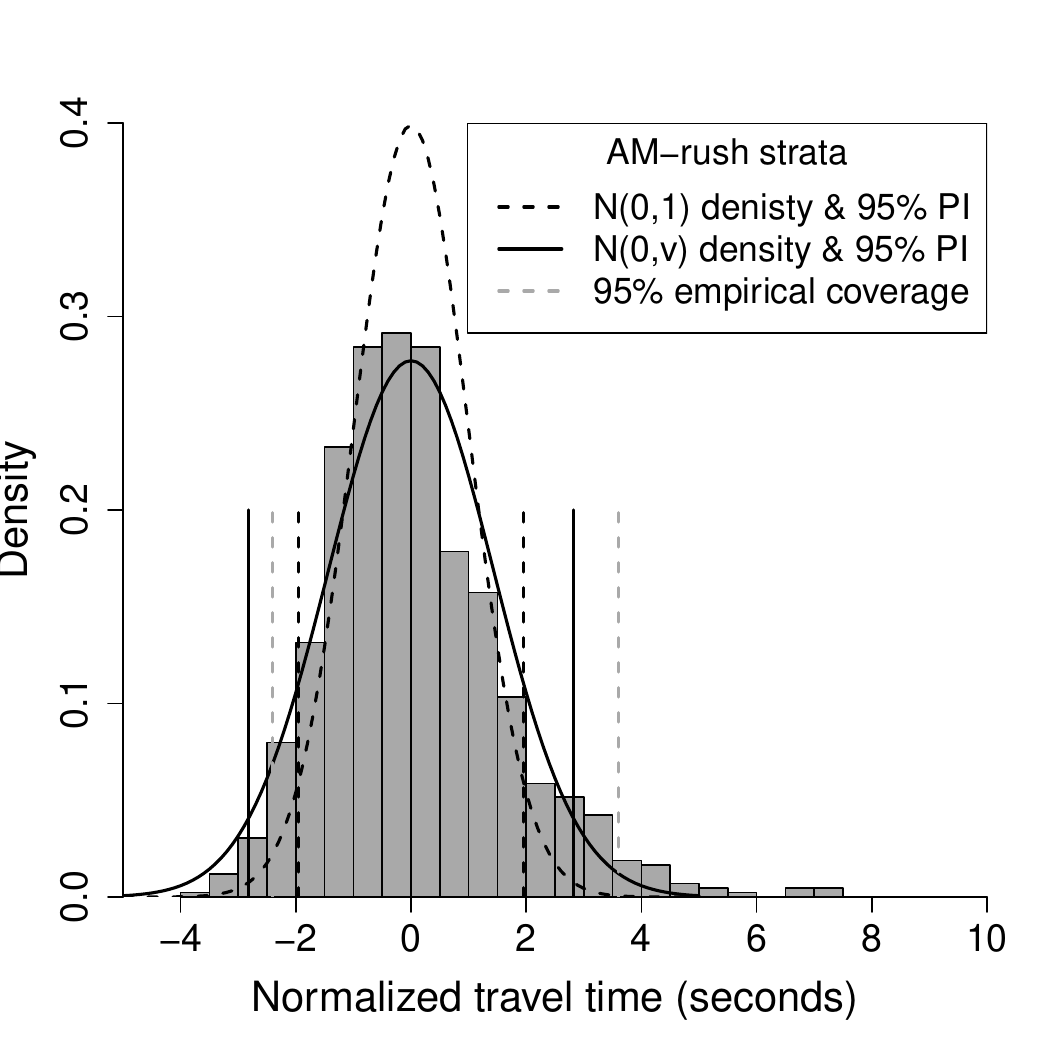}}
  \subfloat[][]{\includegraphics[width=0.5\textwidth]{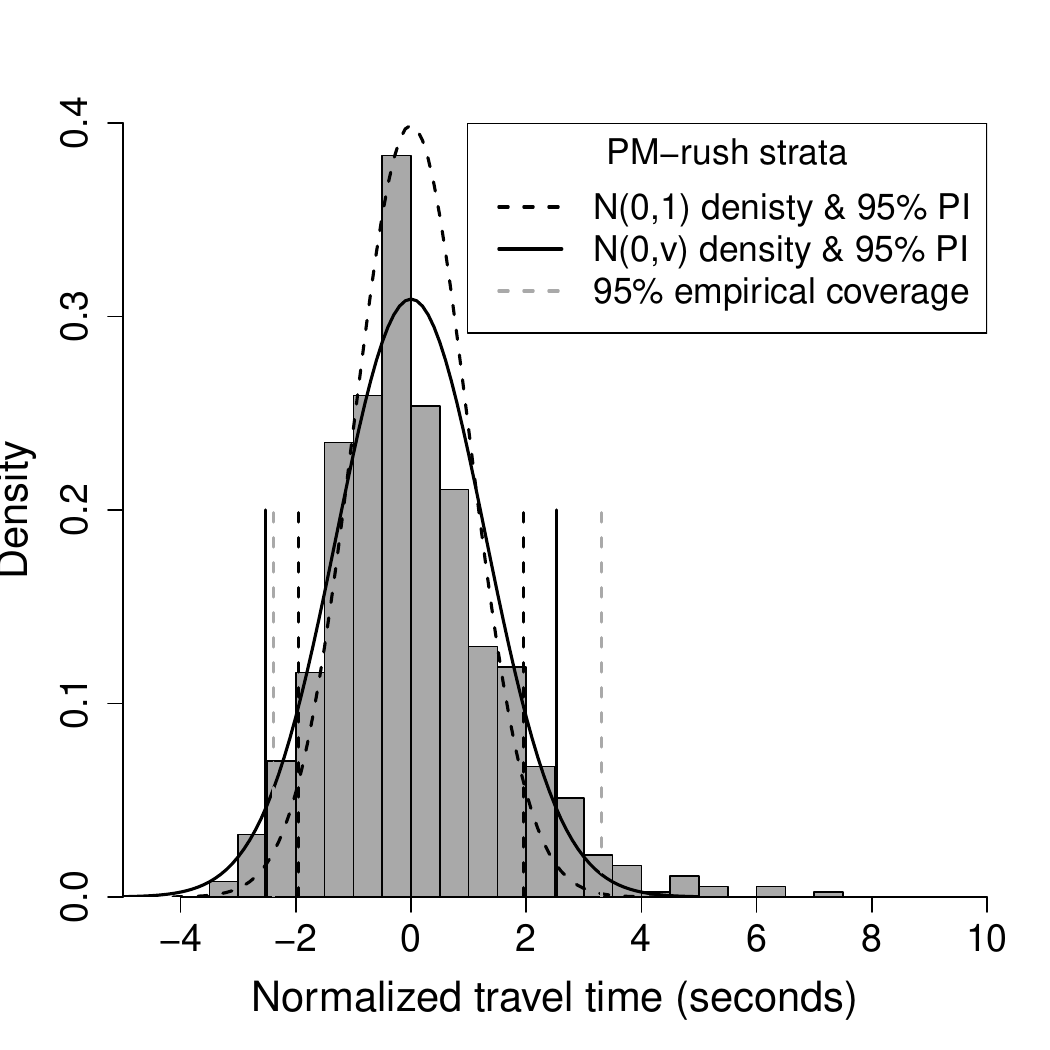}}
  
  \subfloat[][]{\includegraphics[width=0.5\textwidth]{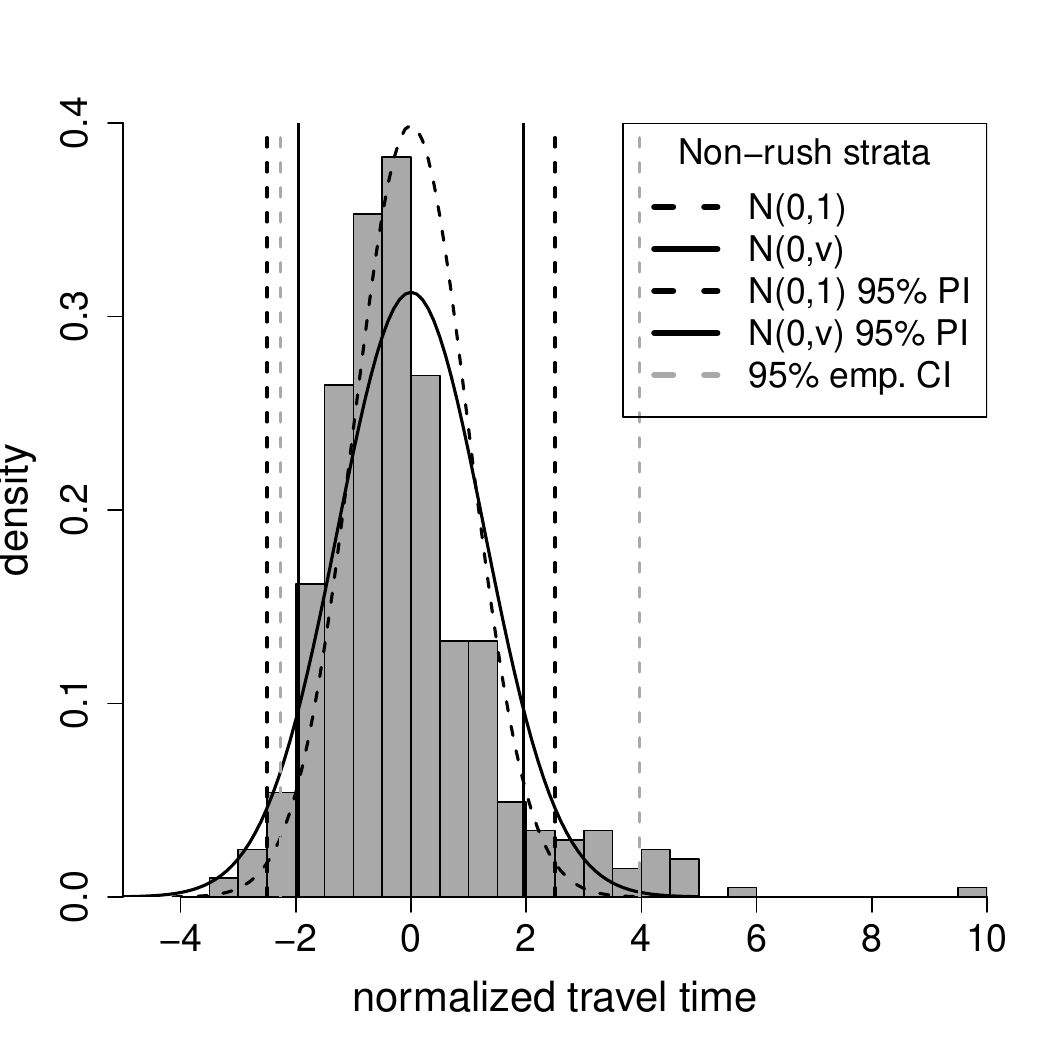}}
  \subfloat[][]{\includegraphics[width=0.5\textwidth]{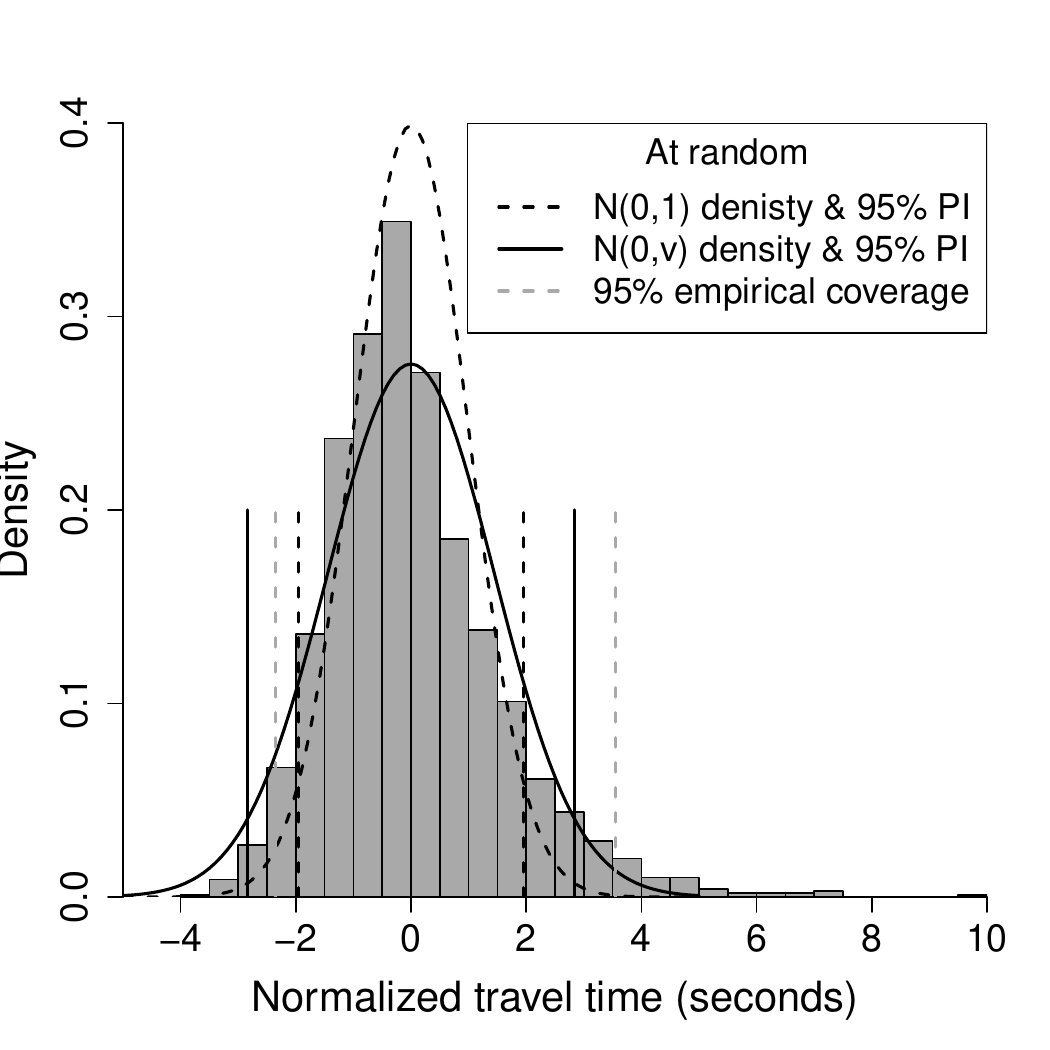}}
  \caption{Histogram of normalized trips (as \(\hatsd_{\path}^{-1}(t_0)(\T_{\path} - \hatmu_{\path}(t_0))\)) from the test set under different sampling methods, with a \(N(0,1)\) density depicted in dashed black, \(N(0, \hatnu)\) in solid black; 95\% prediction intervals are in vertical lines in accordance with density line; in vertical dashed grey is the 95\% empirical coverage intervals.}
  \label{app:fig:trip-hist-prediction-sequence}
\end{figure}

\clearpage

\section{Technical Lemmas}
The proof of the main result in this paper builds on the literature of dynamical systems and Birkhoff's Ergodic Theorem.  In this section, we state the general technical lemmas and definitions that are needed. 

We use \((X, \borelset, \mu)\) to refer to a probability space associated with a random variable \(X\) having a \(\sigma\)-finite Borel algebra \(\borelset\) and a probability measure \(\mu\), such that \(\mu(X)=1\).

\begin{definition} A measure-preserving system (or a dynamical system) is the quadruple \((X, \borelset, \mu, T)\), where \((X, \borelset, \mu)\) is a probability space, and \(T:X\to X\) is a measure-preserving map such \(T^{{-1}}A \in \borelset\) and \(\mu(T^{{-1}}A) = \mu(A)\) for all \(A \in \borelset\); that is \(T\) is \(\mu\)-measurable and \(\mu\)-invariant.
\end{definition}

\(T^{{-1}}\) is the inverse of \(T\). A series of measure-preserving transformations define an {\it orbit}  around a initial point \(x_{o} \in X\), as
\[ \{x_{0}, Tx_{o}, T^{2}x_{0}, \dots, T^{n}x_{0} = T \circ T \circ \cdots \circ Tx_{0}\}.\]

\begin{definition}\label{def:equidistributing} Let \((X, \borelset, \mu, T)\) be a measure-preserving system, and let \(A \in \borelset\). We say the orbit \((T^{n}x_{0})_{n\geq 0}\) equidistributes in \(A\) if
  \[ \lim_{N\to \infty} \frac{1}{N}\#\{n \in \{0, 1, \dots, N-1\}: T^{n}x_{0} \in A\} \to \mu(A)\; a.s .\]
Further, we say \(T\) is {\bf equidistributing}, if for every \(A \in \borelset\) the orbit \((T^{n}x_{0})_{n\geq 0}\) equidistributes in \(A\) for almost every \(x_{0} \in X\).
\end{definition}
In a sense, the frequency distribution of the number of visits to \(A\) converges to \(\mu(A)\) almost surely.

\begin{definition}
  \label{def:ergodic} A measure-preserving system \((X, \borelset, \mu, T)\) is called {\bf ergodic}, if for any \(A\in \borelset\) such that \(T^{-1}A = A\), implies that \(\mu(A)=0\) or \(\mu(A)=1\).
\end{definition}

\begin{definition}\label{def:mixing-dynamical} A measure-preserving system \((X, \borelset, \mu, T)\) is called {\bf mixing}, if for any \(A, B\in \borelset\)
  \[ \lim_{n \to \infty} \mu(A \cap T^{-n}B) - \mu(A)\mu(B)=0.\]
\end{definition}

\begin{lemma}[Thm 1.3 \cite{billingsley1965ergodic}] \label{lem:ergodic-thm} On a probability space \((X, \mathcal B, \mu)\), let \(T: X \to X\) be a measure-preserving transformation. If a function \(f\) is \(L^{1}(X, \mathcal B, \mu)\), then there exists an \(L^{1}(X, \mathcal B, \mu)\) invariant function \(g\) such that \(\int g d\mu = \int fd\mu\), and
  \begin{equation} 
    \lim_{n \to \infty} \frac{1}{n} \sum_{k=0}^{n-1}f(T^{k}x_{o}) = g(x_{o})\quad a.e \text{ (almost everywhere).}
  \end{equation}
  If the system is ergodic, i.e.  \(T\) is equidistributing, then \(g(x_{o}) = \int f d \mu\) a.e.
\end{lemma}

Essentially, ergodicity entails that the system tends to forget the initial value \(x_{0}\). Lemma \ref{lem:ergodic-thm} is an adaptation of \cite[Thm 1.3]{billingsley1965ergodic}, thus will not be proven. To prove our results, we need the following series of lemmas and examples.

\begin{example}[Prop. 2.16 \cite{einsiedler2013ergodic}]\label{ex:rotation} Let \(([0,1], \borelset ([0,1]), \lambda)\) be the \([0,1]\) metric space equipped with the Lebesgue measure \(\lambda\). Let \(Tx = T(x) \pmod 1 = x + \alpha \pmod 1\). Then, if \(\alpha \in  \R\setminus \Q\)(irrationals) the system is ergodic, if \(\alpha \not \in \mathbb Q\), the system is not ergodic.
\end{example}

Part of our results require the ergodicity and mixing of random rotation dynamical systems. The following lemma establishes the ergodicity results with proof in Section \ref{proof:technical-lemma-ergodic-mapping}.
\begin{lemma}  \label{lem:ergodic-mapping} Let \(([0,1], \borelset ([0,1]), \lambda)\) be the \([0,1]\) metric space equipped with the Lebesgue measure \(\lambda\). Let \(T_{k}x = T_{k}(x) \pmod 1 = x + u_{k} \pmod 1\), for \(u_{k} \stackrel{i.i.d}{\sim}\text{Uniform}[0,1]\), then, \(([0,1], \borelset ([0,1]), \lambda, (T_{k})_{k})\) is ergodic. 
\end{lemma}

We show that random rotations are also mixing in the following Lemma with proof in Section \ref{def:proof-random-rotation-mixing}
\begin{lemma}  \label{lem:random-rotations-mixing} Under the setting of Lemma \ref{lem:ergodic-mapping}, random rotations are mixing in the sense of Definition \ref{def:mixing-dynamical}.
\end{lemma}

\begin{lemma}[Random rotations are random variables]\label{lem:rand-rotation-variables} Under the setting of Lemma \ref{lem:ergodic-mapping}, for any \(x \in [0,1]\), the family \((T^{k}x)_{k>1} \stackrel{d}{=} (U_{k})_{k>1}\), where \((U_{k})_{k>1} \stackrel{iid}{\sim}\)Uniform[0,1]. Moreover, for any function, \(f: [0,1] \mapsto \R\), \(f \in L^{2}(\lambda)\) with \(\int fd\lambda = 0\), for any \(x \in [0,1]\), there exists a random variable \(X_{k} \in \R\), such that \( X_{k} \stackrel{a.s.}{=} f(U_{k})\) for all \(k\), with \(\E X_{k} = 0\).
\end{lemma}
\begin{proof} Irrationals are dense in \(\R\), hence an absolutely continuous random variable is almost surely irrational. By Lemmas \ref{lem:ergodic-mapping} and \ref{lem:random-rotations-mixing} \(T\) is ergodic and mixing. By construction and \citet[Thm. 5.10]{kallenberg2006foundations}, for any \(x \in [0,1]\), the family \((T^{k}x)_{k>1} \stackrel{d}{=} (U_{k})_{k>1}\), where \(U_{k} \stackrel{i.i.d}{\sim}\text{Uniform}[0, 1]\). Since \(f \in L^{2}(\lambda)\), by \citet[Thm. 5.11]{kallenberg2006foundations}, there exists a random variable \(X_{k} \in \R\), such that \( X_{k} \stackrel{a.s.}{=} f(U_{k})\) for all \(k\), with \(\E X_{k} = 0\).
\end{proof}
we extend Lemma \ref{lem:rand-rotation-variables} to mixing random variables in the following Lemma.

\begin{lemma}\label{lem:rand-rotation-variables-mixing} Under the setting of Lemma \ref{lem:ergodic-mapping}, define \(T_{k}x = x + u_{k}\pmod 1\), where \(u_{k}\sim\) are identically distributed \(\alpha\)-mixing Uniform[0,1] random variables. Then for  any function, \(f: [0,1] \mapsto \R\), \(f \in L^{2}(\lambda)\) with \(\int fd\lambda = 0\), for any \(x \in [0,1]\), \((f(T_{k}x))_{k>1}\) are a sequence or \(\alpha\)-mixing random variables.
\end{lemma}
\begin{proof} Since \(f \in L^{2}(\lambda)\), it is measurable. By the transfer probability argument in \citet[Thm. 5.10 \& 5.11]{kallenberg2006foundations}, for any \(x \in [0,1]\), for every \(k\), there exists a random variable \(X_{k}\stackrel{a.s.}{=}f(T_{k}x) \stackrel{a.s.}{=}f(U_k)\), for some \(U_{k}\sim\)Uniform[0,1]. Hence for any \(A_{k} \in \sigma(X_{k})\), the \(\sigma\)-algebra generated by \(X_{k}\), and \(T_{k}x = T_{k-1}x + u_{k} \pmod 1\), \(u_{k} \in [0,1]\), we have
  \begin{align*}
    \P(\{X_{1}\in A_{1}\}\cap \{X_{n}\in A_{n}\}) &= \P(\{u_{k}: f(T_{1}x) \in A_{1}\} \cap \{u_{n}: f(T_{n}x) \in A_{n}\}) \\
                                                  & = \P(\{u_{k} : T_{1}x \in f^{-1}(A_{1})\} \cap \{u_{n} : T_{n}x \in f^{-1}(A_{n})\}) \\
                                                  & = \P(\{u_{k} : T_{1}x \in \bar A_{1}\} \cap \{u_{n} : T_{n}x \in \bar A_{n}\}) \\
                                                  & = \P(\{U_{1} \in \bar A_{1}\} \cap \{U_{k}\ \in \bar A_{n}\}),
  \end{align*}
  where \(\bar A_{k} = f^{-1}(A_{k}) \in \sigma (T_{k}x)\). Hence, if the right-hand side is mixing, so is the left-hand side, and vice versa.
\end{proof}

Our proof of Theorem \ref{thm:predictive-traveltime} builds on the following lemma on central limit theorem for random rotation maps.
\begin{lemma}[CLT for random rotations]\label{lem:CLT-random-rotation} Under the setting of Lemma \ref{lem:ergodic-mapping},  for any function, \(f: [0,1] \mapsto \R\), \(f \in L^{2}(\lambda)\) with \(\int fd\lambda = 0\), then for any \(x \in [0,1]\), we have
   \[\frac{1}{\sqrt{n}}\sum_{k=1}^{n}f(T^{k}x) \xrightarrow d N \bigg (0, \int f^{2}d\lambda \bigg )\quad \text{as } n \to \infty.\]
 \end{lemma}
 \begin{proof} From Lemma \eqref{lem:rand-rotation-variables} we know that there exists a random variable \(X_{k} \in \R\), such that \( X_{k} \stackrel{a.s.}{=} f(U_{k})\) for all \(k\), with \(\E X_{k} = 0\). Hence, by classical central limit theorem \citet[Prop. 4.9]{kallenberg2006foundations}, we have
   \[ \frac{1}{\sqrt{n}}\sum_{i=1}^{n}f(T^{k}x)  \xrightarrow d \frac{1}{\sqrt{n}}\sum_{i=1}^{n}X_{k}  \stackrel{d}{=}  N(0, \E X_{1}^{2}), \quad \text{as } n \to \infty.\]
 \end{proof}

We state standard results from \cite{berbee1987convergence}.
\begin{lemma}[Thm. 1.2 \cite{berbee1987convergence}] \label{lem:berbee} Suppose \((X_{n}, n \geq 0)\) is a sequence of \(\alpha\)-mixing bounded random variables with zero mean. If \(\sum_{n>0}n^{-1}\alpha(n) < \infty\), then \(n^{-1}\sum_{n>0}X_{n} \to 0\) a.s.
\end{lemma}

\section{Proof of Lemma \ref{thm:slln-uniform}}\label{app:proof-lemma-slln-stationary}

Our proof for Lemma \ref{thm:slln-uniform}, and the main part of our analysis, is organized as follows:
\begin{enumerate}
\item  Lemma \ref{lem:ergodic-mapping} established that random rotations of the form \(Tx = x + u \pmod1\), \(u \sim \) Uniform[0,1] are equidistributing and hence ergodic in the sense of Definition \ref{def:ergodic} 
\item Lemma \ref{lem:random-rotations-mixing} established that random rotations are also mixing in the sense of \eqref{def:mixing-dynamical}. In general, irrational rotations are ergodic but not mixing.
\item Lemma \ref{lem:rand-rotation-variables} established that random rotation dynamical systems are equal in distribution to a random variable. 
\item By the supremum part in the defining of \(\alpha\)-mixing in \ref{def:alpha-mixing}, we know that \(\alpha\)-mixing systems are also mixing.
\item By Lemma \ref{lem:rand-rotation-variables-mixing} we know that random rotation dynamical systems generated by \(\alpha\)-mixing sequence of random variables are equal in distribution to some \(\alpha\)-mixing sequence of random variables.
\item Lastly, since our established dynamical system is both \(\alpha\)-mixing and random, we utilize direct probabilistic results for mixing sequences to establish a strong law of large numbers for travel time.
\end{enumerate}

\begin{proof}[Proof of Manuscript Lemma \ref{thm:slln-uniform}] Our first condition is that \(\path\) is a random walk on \(G\). Without loss of generality, we will assume that every edge \(e\) has a unit length (i.e.~\((d_{e}=1, e \in E)\)), thus, travel time becomes
\begin{equation}\label{eq:travel-time-1}
  \T_{\path} =  \sum_{e \in \rho} m_{e}(\tau) + \sum_{e \in \rho} \epsilon_{e}(\tau).
\end{equation}

Example \ref{ex:rotation} defined \(\alpha\) as a constant, while in \eqref{eq:recursive-map} it is the random variable \(d_{e}S_{e}(t_{e})\). Hence, by construction and Lemmas \ref{lem:random-rotations-mixing} and \ref{lem:rand-rotation-variables}, we have that \((\epsilon_{e}(\tau), e\in \path )\) is an \(\alpha\)-mixing dynamical system with random rotations. They are \(\alpha\)-mixing since the sequence \((U_{e}, e \in \path)\) is an \(\alpha\)-mixing sequence of Uniform[0,1] random variables. By Definition \ref{def:travel-time} and Lemma \ref{lem:berbee}, we have \(n^{-1} \sum_{e \in \path}\epsilon_{e}(\tau) \xrightarrow {a.s.} 0\).

It remains to be shown that \(n^{-1}\sum_{e \in \rho} m_{e}(\tau)\) converges to a constant that is independent of initial conditions. By Definition \ref{def:travel-time}, we know that \(G\) has a finite node set \(N\), which we denote by \(G_{N}\). By construction, transportation networks have bounded degrees \(\sup_{e \in E}\text{deg}(e) <C_{1}\) for some \(C_{1}< \infty\). From Polya's Theorem on recurrence of random walks in the plane (see \citet[Sec. 2.14]{doyle1984random} and \citet[Thm. 1.1, Cor. 1.2]{benjamini2011recurrence}), \(G\) is recurrent with probability 1 (\(G_{N}\) is a 2-dimensional planar graph).

  Our transportation network \(G\) is equipped with random bounded weights \((S_{e}, e \in E)\). Since \(G\) is finite and \((S_{e}, e \in E)\) are bounded away from 0, then the transportation network is also recurrent with probability 1, meaning that an arbitrary long trip would return to the starting edge with probability 1.

  For each \(e \in E\), let \((\tau_{i}(e))_{i}\) be the almost sure recurrent random times of \(e\). By the recurrence property, we have \(\tau_{i}(e) < \infty\) a.s. for all \(i \in \Z\). Define \(Z_{i}(e) = \tau_{i}(e) - \tau_{i-1}(e)\) for \(i>1\), and \(Z_{1}(e) = \tau_{1}(e)-t_{0}\), the recurrence time difference. By the stationarity of \(G\), \((Z_{i}(e))_{i}\) are independent stationary random variables.

  We first treat each edge \(e \in E\) separately, and show that
  \[\frac{1}{n_{e}} \sum_{i=1}^{n_{e}} m_{e}(\tau_{i}) \to \mu_{e},\]
  where \(\mu_{e}\) is a constant that is independent of recurrence times \((\tau_{i}(e))_{i}\), and \(n_{e}\) is the count of the latter. By the continuity of \(m_{e}(t)\), it is Lebesgue measurable (\(\lambda\)-measurable). Let \(a\)  be the length of the seasonality cycle of \(m_{e}(t)\). Then \(m_{e}\) is \(L^{1}([0,a], \mathcal B[0,a], \lambda)\). 

  By Example~\ref{ex:rotation} and Lemma \ref{lem:ergodic-mapping}, \((\tau_{i}(e))_{i}\) define an equidistributing rotation mapping on the circle \([0,a]\),  with initial point \(x_{0} = t_{0} + Z_{1}(e) \pmod a\), such that
 \[(\tau_{i}(e))_{i} = \bigg ( x_{0}, Tx_{0} = x_{0} +  Z_{2}(e) \pmod a ,T^{2}x = Tx_{0} + Z_{3}(e) \pmod a, ... \bigg ). \]
 Then
 \begin{equation} \label{eq:edge-convergence}
   \frac{1}{n_{e}} \sum_{i=1}^{n_{e}} m_{e}(\tau_{i}) = \frac{1}{n_{e}} \sum_{i=1}^{n_{e}} m_{e}(T^{i}x_{0}) \stackrel{n_{e}}{\longrightarrow} \frac{1}{a}\int_{0}^{a}m_{e}(\lambda)d\lambda \quad  (a.e.)
   \end{equation}
   By Theorem \ref{lem:ergodic-thm}, \(\int_{0}^{a}m_{e}(\lambda)d \lambda\) is independent of initial conditions, the \( \frac{1}{a}\) is to convert the integral to a probability. It is easy to see that
 \[  \frac{1}{a}\int_{0}^{a}m_{e}(\lambda)d\lambda = \E[m_{e}] = \E[\E[S_{e}(t)]] =\E[S_{e}] = \mu_{e},\]
the unconditional expected speed. The Tower's property was used since under stationarity, \(t\) is an index of sub-\(\sigma\)-algebras \(\F_t \subset \F\), where \(\F\) is the space of events of \(S\). This is not surprising, since ergodic dynamical systems have the property that space averaging equals time averaging (the sum in \eqref{eq:edge-convergence}). Combining our results, we have 

\begin{equation}
  \label{eq:m-convergence}
  \frac{1}{n}\sum_{e\in \path}m_{e}(\tau) =\sum_{e \in E}\frac{n_{e}}{n}\frac{1}{n_{e}}  \sum_{i=1}^{n_{e}}m_{e}(\tau_{i}) \stackrel{n}{\longrightarrow} \sum_{e \in E} \spi_{e}  \mu_{e} = \mu, \quad a.s.
\end{equation}
By Empirical process theory we have \(n_{e}/n \xrightarrow {a.s.} \spi_{e} \in [0,1]\) a constant such that \(\sum_{e \in E}\spi_{e}=1\), hence, \(\mu\) is the invariant expected travel time of an edge over the map \(G\).

If \(\path\) is a simple random walk, then \(\spi_{e}\) would be proportional to the degree distribution of \(e\),  otherwise proportional to the weights assigned to \(e\).  We conclude the proof of Lemma \ref{thm:slln-uniform}.
 \end{proof}
 
\subsection{Proof of Lemma \ref{lem:ergodic-mapping}}\label{proof:technical-lemma-ergodic-mapping}
We first require a probabilistic version of Example \ref{ex:rotation}, which can be deduced by the recent results of \cite{limic2018equidistribution}. A general family of maps (not necessarily random) \(\bT:[0,1] \mapsto [0,1]\), where \(\bT = (T_{k})_{k}, T_{k}[0,1]\mapsto [0,1]\), is sufficiently mixing to be equidistributing, if and only if the {\it Weyl criterion} \citep{weyl1916gleichverteilung} \(W_{N}(\bT, m)\) goes to 0 (Lebesgue-a.s.) as \(N \to \infty\), where 
\begin{equation}\label{eq:weyl}
  W_{N}(\bT, m) = \frac{1}{N} \sum_{k=1}^{N} \exp(2\pi i m T_{k} ),
\end{equation}
for all \(m \in \Z\setminus \{0\}\). The above characterization comes from a Fourier analysis. In dimension 1, the class of complex exponentials \(x \mapsto \exp(2\pi i m x), m \in \Z\) is orthonormal in \(L^{2}[0,1]\), and by the Stone-Weierstrass theorem, this class is dense in the periodic continuous functions on \([0,1]\) with respect to the sup-norm. This result allows us to establish equidistributing results for probabilistic mapping. \cite{limic2018equidistribution} defined a Wely-like probabilistic criterion by defining the following random variable
\begin{equation}
  \label{eq:random-map}
  Y_{k}(m) = \exp(2\pi i m T_{k} ), 
\end{equation}
for random maps \(\bT = (T_{k})_{k}, T_{k}: [0,1] \mapsto [0,1]\). The following Lemma gives us conditions regarding when a random mapping is equidistributing. 
  \begin{lemma}[Lem 2.2 of \cite{limic2018equidistribution}]\label{lem:equidistribution-conditions} Let \((T_{k})_{k}\) be a sequence of random maps, and \(Y_{k}(m)\) be as in \eqref{eq:random-map}. If for each \(m \in \Z\setminus \{0\}\)
    \begin{equation} \label{eq:equidistribution-conditions}
      |\E Y_{k}(m)\overline Y_{l}(m) + Y_{l}(m)\overline Y_{k}(m) | = O(| k -l | ^{\delta}),
    \end{equation}
    for some \(\delta(m) >0\), then \((T_{k})_{k}\) is completely equidistributed in \([0,1]\).
\end{lemma}

Using Lemma \eqref{lem:equidistribution-conditions}, we show that the random rotation of Example \ref{ex:rotation} indexed by i.i.d uniform random numbers, such that \(T_{k}x = T_{k}(x) \pmod 1 = x + u_{k} \pmod 1\), where \(u_{k} \stackrel{i.i.d}{\sim}\text{Uniform}[0,1]\), is equidistributing.

\begin{proof} By Lemma \ref{lem:ergodic-thm} and Example \ref{ex:rotation}, we know that the system is measure-preserving; to show that it is ergodic, it suffices to satisfy condition \eqref{eq:equidistribution-conditions} of Lemma \ref{lem:equidistribution-conditions}. For any \(x \in [0,1]\), \(T_{1}(x) = x + u_{1}\), and \(T_{k}(x) = T_{1} \circ T_{2} \cdots \circ T_{k}(x) = x + \sum_{i=1}^{k}u_{k}\). For each \(m \in \Z\setminus\{0\}\), let \(s = \sum_{i=l}^{k}u_{i}\), then
  \begin{align*}
    \E[Y_{k}\overline Y_{l} + Y_{l}\overline Y_{k}](m,x) & = \E \bigg [ \exp(2\pi i m \sum_{i=l}^{k}u_{i}) + \exp(-2\pi i m \sum_{i=l} ^{k}u_{i})\bigg]  \\
                                                         & = 2\int_{[0,1]^{{\mid k -l \mid}}} \cos(2\pi m s) ds = 0
  \end{align*}
\end{proof}
This completes the proof of Lemma \ref{lem:ergodic-mapping}.

\subsection{proof of Lemma \ref{lem:random-rotations-mixing}} \label{def:proof-random-rotation-mixing}

To show that the mixing relation of \ref{def:mixing-dynamical} is satisfied for the space \([0,1] \subset \R\), it is enough to show that it is satisfied for dyadic intervals, since unions of dyadic intervals form an algebra and generate all Borel sets on [0,1], or any \(I \subset \R\). Thus, consider the following sets
\[ A = \Bigg [\frac{t}{2^{i}}, \frac{t+1}{2^{i}}  \Bigg ], \quad  B = \Bigg [\frac{s}{2^{j}}, \frac{s+1}{2^{j}}  \Bigg ], \quad i,j \in {\mathbb N}, 0 \leq t < 2^{i}, 0\leq s < 2^{j}.\]

Without loss of generality, we will assume that  \(i < j\). Consider the random rotations \((T_{k})_{k>0}\), where \(T_{k}x = x + u_{k}\pmod 1\), where \(u_{k}\) are \(i.i.d\) Uniform[0,1]. Then for \(i<j\), we can find a \(u_{0} \in [0,1]\) such \(u_{0} + 2^{-j}(s+1) \pmod 1= t 2^{-i}\); in this sense, for the Lebesgue measure \(\lambda\), we have the following 4 regions:

\begin{itemize}
\item for \(u_{0}\leq u < u_{0} + 2^{-j}\), we have \(\lambda(A \cap [B +u \pmod 1]) = u - u_{0}\).
\item for \(u_{0} + 2^{-j} \leq u < u_{0} + 2^{-i}\), we have  \(\lambda(A \cap [B +u \pmod 1]) = 2^{-j}\).
\item for \(u_{0} + 2^{-i} \leq u < u_{0} + 2^{-i}+2^{-j}\), we have  \(\lambda(A \cap [B +u \pmod 1]) = (u_{0} + 2^{-i} + 2^{-j} - u)\).
  \item \(\lambda(A \cap [B +u \pmod 1]) = 0\) otherwise.
\end{itemize}

By change of variables we can assume that \(u_{0} = 0\), then 
\begin{align*}
  \lambda (A \cap T^{-1}B) &= \int_{0}^{2^{-j}}udu +\int_{2^{-j}}^{2^{-i}}\frac{1}{2^{j}}du  + \int_{2^{-i}}^{2^{-i} + 2^{-j}}\Big ( \frac{1}{2^{i}} + \frac{1}{2^{j}} -u \Big ) du \\
   &= \frac{1}{2^{i}}\frac{1}{2^{j}} = \lambda(A)\lambda(B).
\end{align*}

Lastly, by the invariance property of random rotations to the Lebesgue measure, and induction, we let \(\lambda(A\cap T^{-n}B)  = \lambda(A)\lambda(B)\), satisfying Definition \ref{def:mixing-dynamical}. This completes the proof of Lemma \ref{lem:random-rotations-mixing}. 

\section{Proof of Theorem \ref{thm:clt-traveltime-offline}}\label{app:proof-clt-offline-online}
There are a few methods to prove Theorem \ref{thm:clt-traveltime-offline}. A targeted method would require the utilization of the network structure, the random walk and ergodicity of random rotations. This is a whole endeavour in itself, and so for the purpose of brevity, we rely on known central limit results for non-stationary random variables. We relate a random rotation dynamical system to random variables by \ref{lem:rand-rotation-variables}.

Let \(\{X_{ni}, 1 \leq i \leq k_{n}\}\) and \(n \in \Z\) be a triangular array of the random variables \((X_{i}, i \in \Z)\), where \(k_{n} \to \infty\). Let \(\trho_{max   }\) be a dependency measure between any two non-empty subsets \(A, B \subset \{1,2, \dots, k_{n}\}\) of rows of the array that are at least \(k\) distance apart, as
\begin{equation}
  \trho_{\max}(k)=\sup_{k} |\trho(\sigma(X_{ni}, i \in A),  \sigma(X_{ni}, i \in B)\big )|, \quad  \min_{i \in A, j \in B}|i-j| \geq k.
\end{equation}

\begin{lemma}[Coro. 2.1 \cite{peligrad1996asymptotic}] \label{lem:rho-mixing} Let \((X_{i}, i \in \Z)\) be an \(\alpha\)-mixing sequence, with \(\E[X_{i}]=0\) for all \(i\), and  \((X_{i}^{2}, i \in \Z)\)  is a uniformly integrable family. Define the triangular array \(\{a_{ni}X_{i}, 1 \leq i \leq n\}\), for some constants \(\{a_{ni}\}\), and denote \(\sigma^{2}_{n} = \E[(\sum_{i=1}^{n}a_{ni}X_{i})^{2} ]\). Assume that
    \begin{equation} \label{app:eq:conditoin-max}\max_{1 \leq i \leq n}\frac{|a_{ni}|}{\sigma_{n}} \to 0, \quad \text{as } n \to \infty, 
    \end{equation}
    and
    \begin{equation}
    \label{eq:normalized-variance}
    \sup_{n}\sigma_{n}^{-2}\sum_{k=1}^{n}a_{nk}^{2} < \infty.
  \end{equation}
  Assume in addition, that \(\lim_{n\to \infty }\trho_{max}(n) < 1\), then \(\sigma_{n}^{-1}\sum_{i=1}^{n}a_{ni}X_{i} \xrightarrow d N(0,1)\).
  \end{lemma}
  
  \begin{proof}[Proof of Theorem \ref{thm:clt-traveltime-offline}] We would like to show that
    \[\frac{\T_{\path} - n\mu}{\sqrt n} \xrightarrow d N(0, \sigma^{2}).\]

    By Lemma \ref{thm:slln-uniform}, we have that \(n^{-1}\sum_{e \in \path}d_e m_{e}(\tau) \xrightarrow {a.s.} \mu\), where \(\mu\) is a constant. By reverse application of Slutsky's theorem, we have
\begin{align*}
  \lim_{n\to \infty} \frac{\T_{\path} - n\mu}{\sqrt{n}} & = \lim_{n\to \infty}\frac{\T_{\path} - \sum_{e \in \path}d_{e}m_{e}(\tau)}{\sqrt n} 
  =\lim_{n \to \infty} \frac{1}{\sqrt{n}}\sum_{e\in \path}d_{e}\epsilon_{e}(\tau)\\ & = \lim_{n\to \infty}\frac{1}{\sqrt n}\sum_{e \in \path}d_{e}\sigma_{e}(\tau)X_{e}.
\end{align*}
Here \((X_{e},e \in \path)\) is a sequence of \(\alpha\)-mixing random variables, having \(\E[X_{e}]=0\), and \(\E[X_{e}^{2}]=1\) for all \(e \in \path\). \(\sigma_{e}(\tau)\) is the standard deviation of \(\epsilon_{e}(\tau)\), such that \(\Var(\epsilon_{e}(\tau))=\Var(\sigma_{e}(\tau)X_{e}) = \sigma_{e}^{2}\). 
    By Definition \ref{assump:seasonality-of-speed-on-a-road}, \((S_{e}, e \in E)\) is a family of bounded random variables, and thus \((S_{e}^{2}, e \in E)\) are square integrable. Moreover, \(d_{e} \) and \(d_{e}\sigma_{e}(\tau)\) are bounded for all \(e\) and \(\tau\).

    Since \(n^{-1}\sigma^{2}_{\path}(\tau) \xrightarrow {a.s} \sigma^{2} \neq 0\), in Theorem \ref{thm:clt-traveltime-offline}, it is easy to see that conditions \eqref{app:eq:conditoin-max} and \eqref{eq:normalized-variance} are satisfied. The condition \(\lim_{n \to \infty}\rho^{*}(n)<1\) is assumed in Definition \ref{def:travel-time}, and thus the results follow from Lemma \ref{lem:rho-mixing}.

    The fact that \(\mu\) does not depend on initial conditions follows from Lemma \ref{thm:slln-uniform} with proof in Section \ref{app:proof-lemma-slln-stationary}; this due mainly from the equidistributing (ergodicity) property of random rotations \ref{lem:ergodic-mapping}. Results for the second moment follow accordingly.
 \end{proof}

 \section{Proof of Theorem \ref{thm:predictive-traveltime}} \label{app:proof:predictive-distribution}

 We prove Theorem~\ref{thm:predictive-traveltime} through a series of Lemmas. 
 
 \begin{lemma} \label{lem:rotation-by-average} Following \eqref{eq:mu-recursive}, let \((\Tstar_{e}, e \in \path)\) be defined recursively for every subroute \(\langle \dots, e', e\rangle  \in \path \) as
   \begin{equation}
     \label{eq:tstart}
     \Tstar(e) = \Tstar_{e}= \Tstar_{e'} + d_{e'}m_{e'}(\Tstar_{e'}), 
   \end{equation}
with initial value at \(t_{0}\). Then \((\Tstar_{e}, e \in \path)\) are equidistributing.
 \end{lemma}
 \begin{proof} See Section  \ref{lem:rotation-average}  \end{proof}
 \begin{lemma} \label{lem:rotation-by-average-are-mixing} Following the settings of Lemma \ref{lem:rotation-by-average}, let \(\path\) be a random walk on a transportation network \(G\), and define \((\Tstar_{i}(e), i = 1, \dots, n_{e})\) as the visit times to edge \(e\). Then, the sequence \((\Tstar_{i}(e), i=1, \dots, n_{e})\) is mixing in the sense of Definition \ref{def:mixing-dynamical}.
 \end{lemma}
 \begin{proof} See Section \ref{lem:rotation-average} \end{proof}
 
 \begin{proof}[Proof of Theorem \ref{thm:predictive-traveltime}] Following similar set-up to Theorem \ref{thm:clt-traveltime-offline}, let \(n = |\path|\); we then decompose \eqref{eq:predictive-clt} as
  \begin{equation}
  \frac{\T_{\path} - \mu_{\path}(t_0)}{\sigma_{\path}(t_0)}
    = \frac{\sqrt{n}\sigma}{\sigma_{\path}(\Tstar)} \frac{\T_{\path} - \mu_{\path}(\tau)}{\sqrt{n} \sigma} -  \frac{\sqrt{n}\sigma}{\sigma_{\path}(t_0)}\frac{\mu_{\path}(t_0) - \mu_{\path}(\tau)}{\sqrt{n}\sigma}= I  - II.
  \end{equation}

  By Theorem \ref{thm:clt-traveltime-offline} and Slutsky's theorem, we have
\begin{equation}
  \label{eq:I}
I = \frac{\sqrt{n}\sigma}{\sigma_{\path}(t_0)} \frac{\T_{\path} - \mu_{\path}(\tau)}{\sqrt{n} \sigma} \xrightarrow d \sqrt{\eta}N(0,1), \quad \eta =  \lim_{n \to \infty}\frac{n\sigma^{2}}{\sigma^{2}_{\path}(t_0)}. 
\end{equation}

For \(II\), we know that \(n^{-1}\mu_{\path}(\tau) \to \mu\) a.s.~from Lemma~\ref{thm:slln-uniform}. Hence, we will first show that \(n^{-1}\mu_{\path}(t_0) \xrightarrow {a.s.} \mu\). which requires the deterministic times $\Tstar = (\Tstar_{e}, e \in \path)$ to be equidistributing, in the sense of Definition \ref{def:equidistributing}. This is established by Lemmas \ref{lem:rotation-average} and \ref{lem:ergodic-mapping}. Hence, by a similar argument to \eqref{eq:m-convergence}, we have 
\[ \frac{1}{n}\mu_{\path}(t_0) = \frac{1}{n}\sum_{e \in \path} d_e m_{e}(\Tstar) = \sum_{e \in E} \frac{n_{e}}{n}\frac{1}{n_{e}}\sum_{i=1}^{n_{e}}d_e m_{e}(\Tstar_{i}(e)) \xrightarrow {n} \sum_{e \in E}\pi_{e}\mu_{e}, \quad a.s.\]

As shown in the proof of Lemma \ref{thm:slln-uniform} in Section \ref{app:proof-lemma-slln-stationary}, \(\path\) is a random walk on \(G\) and hence recurrent with probability 1; see~\citet[Sec. 2.14]{doyle1984random} and \citet[Thm. 1.1, Cor. 1.2]{benjamini2011recurrence}.

Assume that the period of each \((m_{e}(t), e \in E)\) is of length \(a>0\). Since \(m_{e}(t) \in L^{2}(\lambda)\), let \((\Tstar_{i}(e), i = 1, \dots n_{e})\) be the visit times to edge \(e\) in \((\Tstar_{e}, e \in \path)\). Then, by Lemma \ref{lem:rotation-by-average-are-mixing}, the mapping \((\Tstar_{i}(e), i = 1, \dots n_{e})\) is mixing, since it can be written as
 \[\Tstar_{i}(e) = \Tstar_{i-1}(e) + U_i \pmod a,\]
 where \((U_{i}, i=1, \dots, n_{e})\) are i.i.d Uniform\([0,a]\) random variables. By Lemma  \ref{lem:CLT-random-rotation}, for every \(e \in E\),
 \[n_{e}^{-1/2}\bigg (\sum_{i=1}^{n_{e}} d_e m_{e}(\Tstar_{i}) - \mu_{e} \bigg )\xrightarrow d N(0, \tilde \sigma_{e}^{2}),\quad \text{as } n_{e} \to \infty,\]
where 
  \[\tilde \sigma_{e}^{2}= d_e^2 \bigg [\int m_{e}^{2}(t)dt - \bigg (\int m_{e}(t)dt\bigg )^{2} \bigg ].\]

  By \eqref{eq:speed-regimes}, conditional on speed-states \(\Pi\), and the property of the sum of independent normal variables, we have that 
  \begin{equation} \label{eq:clt-trip-specific-proof}
\frac{\mu_{\path}(t_0) - n\mu}{\sqrt n} =   \sum_{e \in E} \frac{\sqrt{n_{e}}}{\sqrt{n}} \frac{1}{\sqrt{n_{e}}} \sum_{i=1}^{n_{e}}\bigg ( d_e m_{e}(\Tstar_{i}) - \mu_{e} \bigg ) \xrightarrow d N(0, \tilde \sigma^{2}), \quad \text{as } n \to \infty,
\end{equation}
where, for \(n_{e}/n \xrightarrow {a.s.} \pi_{e}\), 
\[\tilde \sigma^2= \sum_{e \in E} \spi_{e}d_e^2\sigma_{e}^{2} = \sum_{e \in E} \spi_{e} d_e^2\bigg [\int m_{e}^{2}(t)dt - \bigg (\int m_{e}(t)dt\bigg )^{2} \bigg ].\]

By across-trip dependency  \(I \independent II\), and from \eqref{eq:I} and \eqref{eq:clt-trip-specific-proof}, and by a second application of Slutsky's theorem, we have
\[I - II \stackrel{d}{=} \sqrt{\eta}N(0,1) + \sqrt{\eta}N(0, \tilde \sigma^{2}),\quad \text{as } n \to \infty. \]

This completes the proof of Theorem \ref{thm:predictive-traveltime}.
\end{proof}   

 \subsection{Proof of Lemmas \ref{lem:rotation-by-average} and \ref{lem:rotation-by-average-are-mixing}} \label{lem:rotation-average}

 \begin{proof}[Proof of Lemma \ref{lem:rotation-by-average}] This follows directly from Example~\ref{ex:rotation}, since for an arbitrary \(t>0\), \(m_{e}(t)\) is almost surely irrational (by continuity). Hence, \((\Tstar_{e}, e \in \path)\) is an irrational family of maps, thus equidistributing.
 \end{proof}

  \begin{proof}[Proof of Lemma \ref{lem:rotation-by-average-are-mixing}] Following the proof arguments of Theorem \ref{thm:slln-uniform} in Appendix \ref{app:proof-lemma-slln-stationary}, and the fact that \(\path\) is a random walk on $G$, from the recurrence property in 2-dimensional planar graphs, we have \(\Tstar_{i}(e) < \infty\) a.s. for all \(i \in \Z\).

   Without loss of generality, define \(U_{i}(e) = \Tstar_{i}(e) - \Tstar_{i-1}(e)\) for \(i>1\), the recurrence time difference, with \(U_{1}(e) = \Tstar_{1}(e) - t_{0}\), where $t_{0}$ is the start time of the trip. By the stationarity of \(G\), \((U_{i}(e), i =1, \dots, n_{e})\) are independent continuous stationary random variables. The continuity follows from the fact that \(m_{e}(t)\) is continuous.

   By the transfer probability argument in \citet[Thm. 5.10]{kallenberg2006foundations}, \((U_{i}(e), i =1, \dots, n_{e})\) are equal in distribution to some Uniform[0,1] random numbers. By Lemma \ref{lem:random-rotations-mixing}, let \([0,a]\) be the cycle of edge \(e\), then rotation mapping
   \[\Tstar_{i}(e) = \Tstar_{i-1}(e) + U_i(e) \pmod a,\]
   is mixing.
 \end{proof}
This completes the proof.